\documentclass[journal,12pt,onecolumn]{IEEEtran}

\ifCLASSINFOpdf

\else

\fi

\usepackage{graphicx}
\usepackage{latexsym,bm}
\usepackage{amsmath}
\usepackage{amssymb}
\usepackage{amsthm}
\usepackage{epstopdf}
\usepackage{cite}
\usepackage{mathrsfs}
\usepackage{multirow}
\usepackage[margin=1.0in]{geometry}
\usepackage{booktabs}
\usepackage{array}
\usepackage{caption}
\usepackage{pgfplots}
\usepackage{algorithm}
\usepackage{pbox}
\usepackage{epsfig}
\usepackage{algorithmic}
\usepackage{subcaption}
\pgfplotsset{compat=newest} 
\pgfplotsset{plot coordinates/math parser=false}

\newcolumntype{C}{ >{\centering\arraybackslash} m{3cm} }
\newcolumntype{D}{ >{\centering\arraybackslash} m{6cm} }
\newcolumntype{E}{ >{\centering\arraybackslash} m{3cm} }
\newcolumntype{F}{ >{\centering\arraybackslash} m{9cm} }
\newcolumntype{K}{ >{\centering\arraybackslash} m{1.5cm} }
\newcolumntype{L}{ >{\centering\arraybackslash} m{13.5cm} }
\newcolumntype{M}{ >{\centering\arraybackslash} m{15cm} }

\DeclareMathAlphabet{\mathscrbf}{OMS}{mdugm}{b}{n}
\begin{document}
\begin{titlepage}
\begin{center}
\vspace*{-2\baselineskip}
\begin{minipage}[l]{7cm}
\flushleft
\includegraphics[width=2 in]{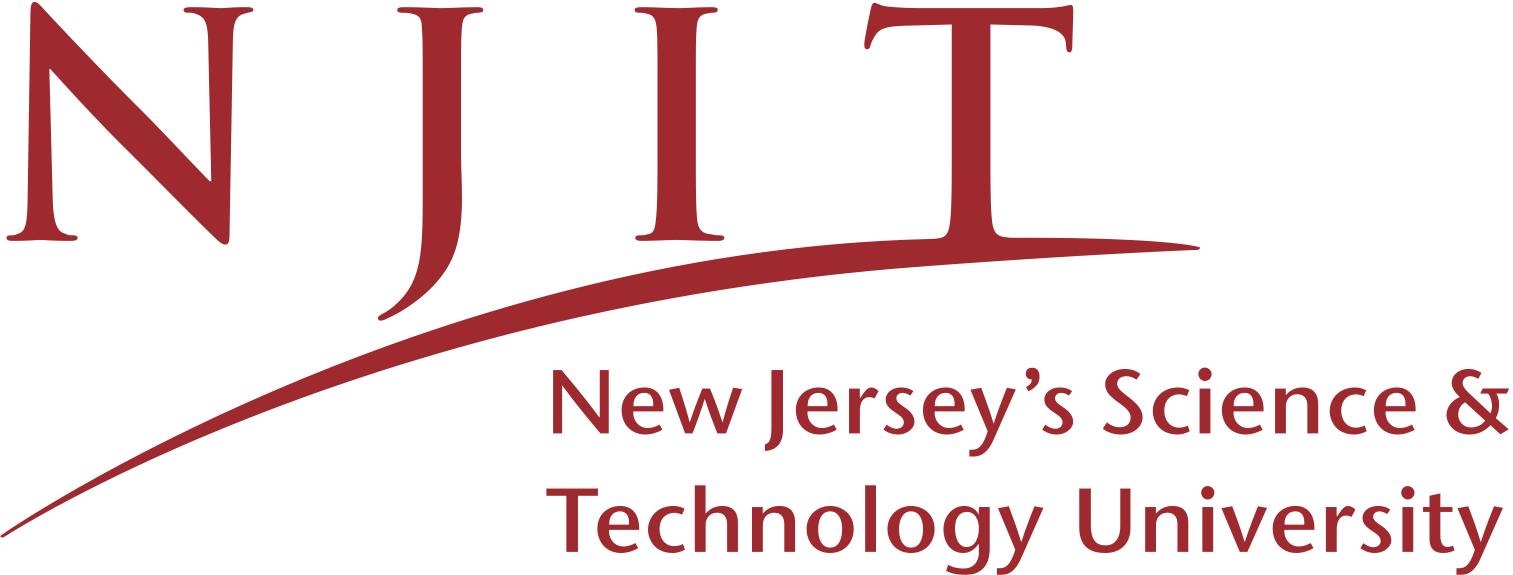}
\end{minipage}
\hfill
\begin{minipage}[r]{7cm}
\flushright
\includegraphics[width=1 in]{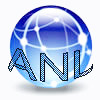}%
\end{minipage}

\vfill

\textsc{\LARGE Optimal Cooperative Power Allocation for \\ [12pt]
Energy Harvesting Enabled Relay Networks}

\vfill
\textsc{\LARGE XUEQING HUANG\\[12pt]
\LARGE NIRWAN ANSARI}\\
\vfill
\textsc{\LARGE TR-ANL-2014-004\\[12pt]
\LARGE May 22, 2014}\\[1.5cm]
\vfill
{ADVANCED NETWORKING LABORATORY\\
 DEPARTMENT OF ELECTRICAL AND COMPUTER ENGINEERING\\
 NEW JERSY INSTITUTE OF TECHNOLOGY}
\end{center}
\end{titlepage}
\begin{abstract}
In this paper, we present a new power allocation scheme for a decode-and-forward (DF) relaying-enhanced cooperative wireless system. While both source and relay nodes may have limited traditional brown power supply or fixed green energy storage, the hybrid source node can also draw power from the surrounding radio frequency (RF) signals. In particular, we assume a deterministic RF energy harvesting (EH) model under which the signals transmitted by the relay serve as the renewable energy source for the source node. The amount of harvested energy is known for a given transmission power of the forwarding signal and channel condition between the source and relay nodes. To maximize the overall throughput while meeting the constraints imposed by the non-sustainable energy sources and the renewable energy source, an optimization problem is formulated and solved. Based on different harvesting efficiency and channel condition, closed form solutions are derived to obtain the optimal source and relay power allocation jointly. It is shown that instead of demanding high on-grid power supply or high green energy availability, the system can achieve compatible or higher throughput by utilizing the harvested energy.
\end{abstract}

\begin{IEEEkeywords}
Power allocation, DF-Relay, Cooperative communications, RF Energy harvesting.
\end{IEEEkeywords}

\IEEEpeerreviewmaketitle
\section{Introduction}
Although current wireless networks still primarily rely on the on-grid or un-rechargeable energy sources, continuous advances in green energy technology has motivated the research of the \emph{green powered wireless network} \cite{6472203}, \cite{Han:2012:ICE}. The concept of energy harvesting (EH) has been proposed to capture and store energy from readily available ambient sources that are free for users, including wind, solar, biomass, hydro, geothermal, tides, and even radio frequency signals \cite{RF}. EH is capable of generating electricity or other energy form, which is renewable and more environmentally friendly than that derived from fossil fuels \cite{RFHARVESTOR}. 

The generic green energy harvesting model adopts the \emph{harvest-store-use} architecture with a storage component (e.g., rechargeable batteries) to hoard the harvested energy for future use. Except for the storage unit, the energy harvester and the energy usage components can be either 1) separated, which allows \emph{simultaneous} energy harvesting and wireless functionality, such as data transmission or reception, or 2) co-located, which adopts \emph{time switching} scheduling between the energy harvesting and consuming processes. Furthermore, the existing literature assumes that the current harvested energy can only flows to latter slots, owing to the \emph{energy half-duplex constraint} \cite{6449245}. So, before performing the wireless functionality, the available residual energy is observable in both architectures, similar to the on-grid powered traditional wireless networks.

It is, however, not trivial to design and optimize the green energy enabled networks owing to the fact that the \emph{energy-arrival rate} of the free energy is determined by the surrounding environment, such as the power generators' geo-locations and weather conditions. Since the energy cannot be consumed before it is harvested, the \emph{opportunistic energy harvesting} results in fluctuating power budget, namely, \emph{energy causality constraint} (EC-constraint). The EC-constraint mandates that, at any time, the total consumed energy should be equal to or less than the total harvested energy, which maybe further limited by the finite battery capacity \cite{5522465}, \cite{energyCausality}.

For the architecture with separated energy harvester and information transmitter, green power management is essential to maximize the system performance while not violating the EC-constraint. Ho and Zhang \cite{6202352} considered the point-to-point wireless system with the energy harvesting transmitter. Optimal energy allocation algorithms are developed to maximize the throughput over a finite time horizon. Similarly, the throughput by a deadline is maximized and the transmission completion time of the communication session is minimized \cite{5441354}, \cite{5992841}. Moreover, the works in \cite{6363767} and \cite{6381384} explored the joint source and relay power allocation over time to maximize the throughput of the three node decode-and-forward (DF) relay system, in which both the source and relay nodes transmit with power drawn from independent energy-harvesting sources.

For the green relay enhanced cooperative wireless network, radio frequency (RF) harvesting is an energy form of particular potential because it enables simultaneous wireless information and power transfer \cite{6489506}. For the relay node (RN) with co-located data and energy reception components, it can either split the received signals between data detector and energy harvester (\emph{power splitting}), or perform the above mentioned two processes sequentially (\emph{time switching}) \cite{6552840}. 

Furthermore, since the half duplex relay is required to transmit and receive over orthogonal time slots \cite{1435648}, the source node (SN) can harness energy from the forwarding signals transmitted by the relay. Inherently, the data transmission and energy harvesting will occur alternatively. So, the co-located time switching architecture can be adopted by SN.


To study the advantage of introducing energy harvesting capability into SN rather than RN, this paper addresses the joint energy management policies for the source and relay nodes, where the DF-relay node is equipped with limited brown or green energy storage, and the source node can harvest energy from the relaying signals. To guarantee a certain level of stability in energy provisioning, a backup non-renewable energy source or complementary green energy source is also available for SN in case the power provided by the RF energy harvester is insufficient. By utilizing the available power thoughtfully, the system throughput is maximized for a given amount of available on-grid power or green power. The derived results can be easily extended to scenarios where both SN and RN can harvest energy from each other's signals.



\section{System Model and Problem Formulation}\label{section1}
\begin{figure*}
\centering
\hspace*{\fill}
                \begin{subfigure}[b]{3 in}
                \includegraphics[width=\textwidth]{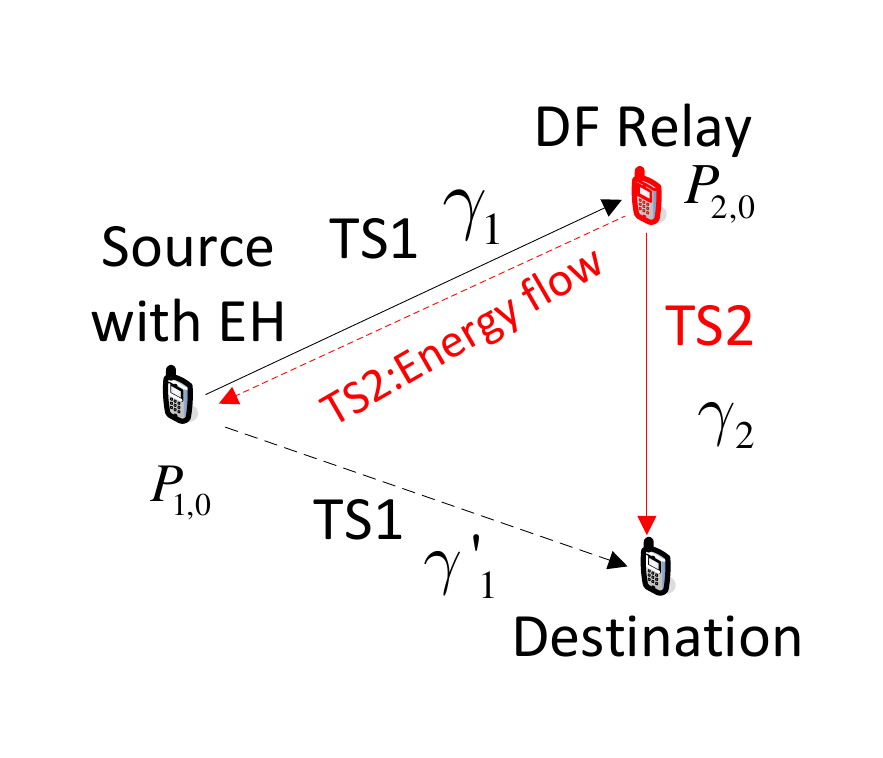}
        \end{subfigure}\hfill
        \begin{subfigure}[b]{3 in}
                \includegraphics[width=\textwidth]{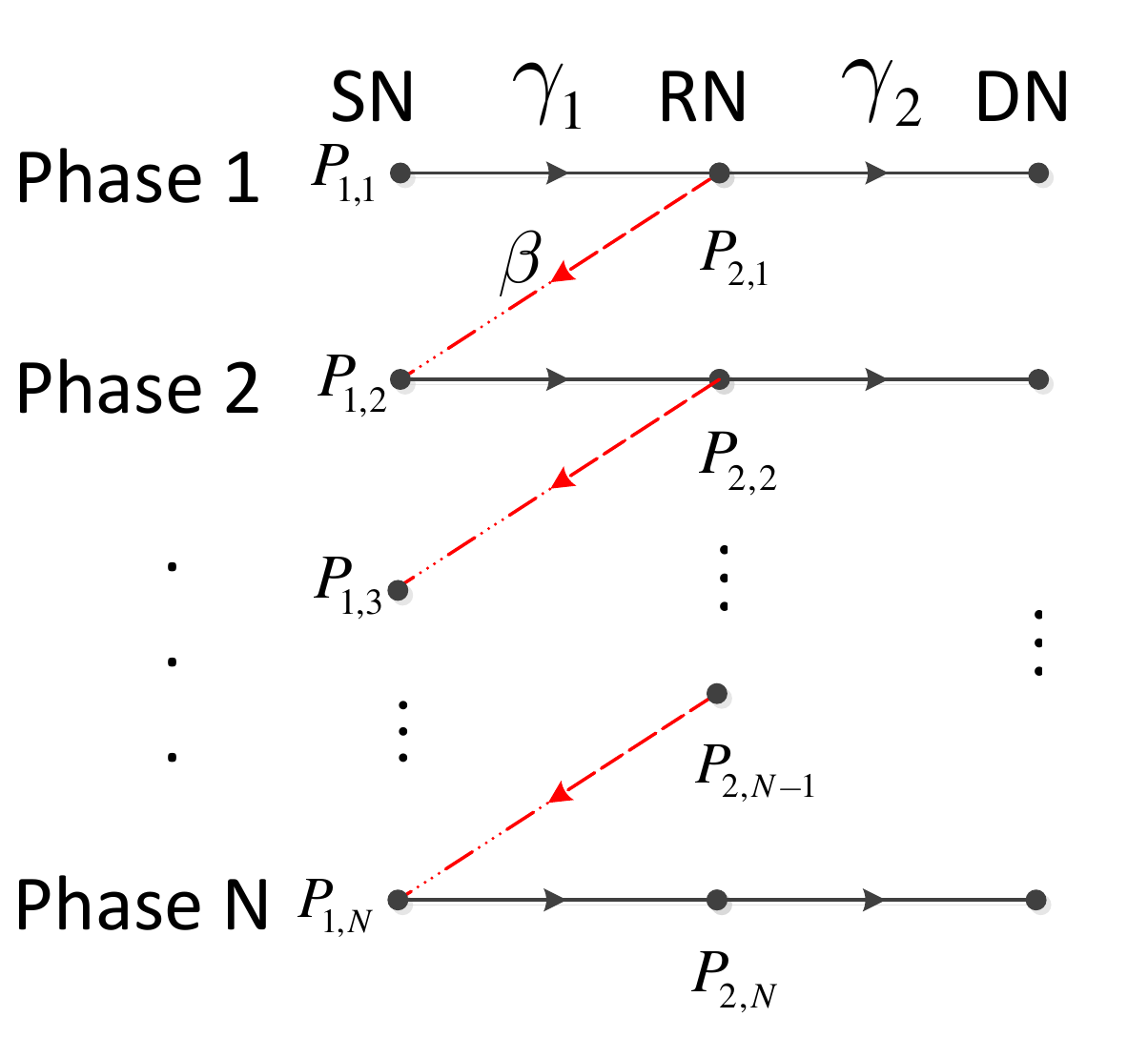}
        \end{subfigure}%
        \hspace*{\fill}
        \caption{Energy and data flows in the DF-relay enhanced system with RF-EH SN}\label{scenario}
\end{figure*}

Consider the Shannon capacity of the half duplex relay system measured over $N$ phases, where $N$ can be the delay requirements of data traffic, and each phase consists of two consecutive time slots (TSs). In each odd TS, SN will transmit data to the relay node, while in the even TS, RN will forward the signal received in previous TS. The amount of the green/brown energy already acquired by SN and RN are $P_{1,0}$ and $P_{2,0}$, respectively, as shown in Fig. \ref{scenario}. The energy harvested by SN in one phase can be used to facilitate future data transmission. 

The total bandwidth occupied by the system is $B$. For the sake of convenience, we assume the constant channel power gains across $N$ phases \cite{6381384}, where $h_{i}$ is the channel gain of the SN-RN link ($i=1$) and the RN-DN link ($i=2$). ${\gamma_{i}} = {{{{\left| {{h_{i}}} \right|}^2}}}/({{{N_0}B}})$ denotes the corresponding normalized signal-noise-ratio (SNR) associated with the channel between SN and RN ($i=1$) as well as that associated with the channel between RN and DN ($i=2$). ${N_0}B$ represents the power of additive white Gaussian noise. Without loss of generality, for now, we assume no direct link exists between SN and DN, i.e., the corresponding SNR $\gamma_1'=0$. The case with direct link will be discussed in the next section.

The goal is to design the optimal power allocation $P_{i,j}$, $i = \{1,2\},j\in\{1,\cdots,N\}$ such that the overall system throughput cross $N$ phases is maximized.

\begin{equation}
\label{initial}
\begin{array}{l}
{C^*} = \mathop {\max }\limits_{{P_{i,j}}} C = \frac{B}{2}\sum\limits_{j = 1}^N {\mathop {\min }\limits_{i = 1,2} \{ \log (1 + {P_{i,j}}{\gamma _i})} \} \\
\begin{array}{*{20}{l}}
{s.t.}&\text{Energy-causility Constaint } EC^*_{j}:&\sum\limits_{k = 1}^j {{P_{1,k}}}  \le {P_{1,0}} + \beta \sum\limits_{k = 1}^{j - 1} {{P_{2,k}}}\\
&\text{Power Budget Constraint }{C^*_1}:&\sum\limits_{j = 1}^N {{P_{2,j}}}  \le {P_{2,0}} \\
&\text{Non-negative Constraint }{C^*_2}:&{P_{i,j}} \ge 0,i\in\{1,2\},j\in\{1,\cdots,N\}
\end{array}
\end{array}
\end{equation}
where $EC_j^*$, $j \in \{ 1, \cdots ,N\}$ is the energy causality constraint of the $j$-th phase. $C_1^*$, represents the budget of the transmission power in RN. $C_2^*$, represents the non-negative power allocation. ${B}/{2}$ is attributed to the half-duplex of the relay channel. $\beta{P_{2,j}}$ is the amount of power harvested in phase $j$ and used in phases after $j$. $\beta={{\eta}} {{\left| {{h_1}} \right|}^2}$ with $\eta$ denoting the energy harvesting efficiency factor \cite{6702854}.


\section{Power Allocation Analysis}\label{section2}
To reveal some insights of the optimal solution, the following propositions are presented.

\emph{Proposition 1}: To maximize the throughput, the power budget constraint ${C^*_1}$ in Eq. (\ref{initial}) is satisfied with equality.
\begin{equation}
\mathop \sum \limits_{j = 1}^N {P_{2,j}} = {P_{2,0}},N \ge 2
\end{equation}
\begin{proof}Define the \emph{\bf{residual power}} of SN and RN at the beginning of the $j$-th phase as $\overline {{P_{1,j}}} $ and $\overline {{P_{2,j}}} $, respectively. Then in the last phase, at least one of the nodes (SN or RN) will use all of its residual energy. $\mathop \sum \nolimits_{j = 1}^N {P_{2,j}} < {P_{2,0}}$ implies ${P_{2,N}} < \overline {{P_{2,N}}} $ and ${P_{1,N}} = \overline {{P_{1,N}}} $.

a) If $\overline {{P_{1,N}}} {\gamma _1} \ge \overline {{P_{2,N}}} {\gamma _2}$, then obviously ${P_{2,N}} < \overline {{P_{2,N}}} $ is not the optimal solution. 

b) If $\overline {{P_{1,N}}} {\gamma _1} < \overline {{P_{2,N}}} {\gamma _2},$ then, there exists a positive ${\alpha _{N - 1}}$ such that 
\[\left( {\overline {{P_{1,N}}}  + {\alpha _{N - 1}}\beta } \right){\gamma _1} = (\overline {{P_{2,N}}}  - {\alpha _{N - 1}}){\gamma _2}\]
where ${\alpha _{N - 1}}= \frac{{\overline {{P_{2,N}}} {\gamma _2} - \overline {{P_{1,N}}} {\gamma _1}}}{{\beta \gamma_1+\gamma_2 }}$.

This means at phase $N$-1, RN will increase ${P_{2,N - 1}}$ by ${\alpha _{N - 1}}$ such that ${P_{1,N}^*} =  {\overline {{P_{1,N}}}  + {\alpha _{N - 1}}\beta }$, ${P_{2,N}^*} = \overline {{P_{2,N}}}  - {\alpha _{N - 1}}$. Since $\mathop {{\rm{min}}}\limits_{i = 1,2} \left\{ {{P_{i,N}^*}{\gamma _i}} \right\} \ge \mathop {{\rm{min}}}\limits_{i = 1,2} \left\{ {{P_{i,N}}{\gamma _i}} \right\} $, the total throughput will increase when ${P_{2,N}^*} = \overline {{P^*_{2,N}}} $.
\end{proof}

\emph{Proposition 2}: In the optimal power allocation, there exists an $\alpha$, $(\alpha\ge0)$, such that the following equality is satisfied.
\[({P_{2,1}}-\alpha){\gamma _2}={P_{1,1}}{\gamma _1}, P_{2,j} {\gamma _2}=P_{1,j} {\gamma _1},j\in\{2,\cdots,N\}\]
\begin{proof}
Since the throughput of phase $j$ is determined by $\mathop {{\rm{min}}}\limits_{i = 1,2} \left\{ {{P_{i,j}}{\gamma _i}} \right\}$, and SN can harvest energy from the signals transmitted by RN, it is reasonable to assume ${P_{1,j}} {\gamma _1}\le{P_{2,j}} {\gamma _2},j\in\{1,\cdots,N\}$. Then, at each phase, we can divide ${P_{2,j}}$ into two parts as illustrated below
\begin{equation}
\label{reas}
\left\{\begin{array}{l}
{P_{2,j}}= {p_j}+{\alpha _j},j \in \left\{ {1, \cdots ,N} \right\}\\
{{P_{1,j}}{\gamma _1}=p_{j}}{\gamma _2},j \in \left\{ {1, \cdots ,N} \right\}
\end{array}\right.
\end{equation}
where ${P_{1,j}}$ is used for {\bf{data transmission}} and ${p_j}$ is for \emph{\bf{data forwarding}}. The \emph{\bf{power supplement}} ${\alpha _j}$ is provided by RN to increase the energy storage of SN, and the harvested ${\alpha_j}\beta$ will be used by SN for the future data transmission.

For any feasible solution with ${\alpha _j} > 0,j \in \left\{ {2, \cdots ,N } \right\}$, an equivalent solution can always be found with 1) the same power allocation for data transmission and forwarding, i.e., $P{^*_{1,j}} = {P_{1,j}}$, $p{^*_j} = {p_j}$, $\forall j \in \left\{ {1, \cdots ,N} \right\}$. 2) power supplements are aggregated to the first phase, i.e., $\alpha {^*} = \mathop \sum \nolimits_{j = 1}^{N } {\alpha _j}$, $\alpha {^*_j} = 0$, $\forall j \in \left\{ {2, \cdots ,N} \right\}$.
\end{proof}

Adopting the previous propositions, the solution to Eq. (\ref{obj_fin}) must be the solution to Eq. (\ref{initial}). 
\begin{equation}
\label{obj_fin}
\begin{array}{l}
{C^*} = \mathop {\max }\limits_{\{ {p_j},{\alpha}\} } C = \frac{B}{2}\sum\limits_{j = 1}^N {\log (1 + {p_j}{\gamma _2})} \\
\begin{array}{*{20}{l}}
{s.t.}&\text{Energy-causility Constraint }{EC}_1\text{:}&{{p_1} \le {P_{1,0}}\gamma}\\
{}&\text{Energy-causility Constraint }{EC}_j\text{:}&{\sum\limits_{k = 1}^j {{p_k}}  \le {P_{1,0}}\gamma + \beta\gamma ({\alpha} + \sum\limits_{k = 1}^{j - 1} {{p_k}} )}\\
{}&\text{Power Budget Constraint }C1\text{:}&{\sum\limits_{j = 1}^N {{p_j}}  + {\alpha} = {P_{2,0}}}\\
{}&\text{Non-negative Constraint }C2\text{:}&{\alpha},{p_j} \ge 0,\forall j\in\{1,\cdots,N\}\\
\end{array}
\end{array}
\end{equation}
where $\alpha$ is the aggregated power supplement provided by RN in the first phase. ${EC}_j$, $j\in\{2,\cdots,N\}$ is the corresponding EC-constraint, and the SNR ratio $\gamma=\gamma_1/\gamma_2$.


To find the optimal solution to Eq. (\ref{obj_fin}), we first consider the scenario where there is only the constant power budget, i.e., constraint $C_1$. According to the water-filling algorithm \cite{Boyd:2004:CO:993483}, the relaxed optimal solution for this scenario is
\begin{equation}
\label{rel_sol}
{\alpha} = 0,{p_j} = {P_{2,0}}/N,j \in \{ 1, \cdots ,N\}
\end{equation}

To check the feasibility of the relaxed optimal solution for the scenario with all of the energy-causality constraints in Eq. (\ref{obj_fin}), we substitute the value of $\alpha$, $p_j$ of Eq. (\ref{rel_sol}) into $EC_j$, $j\in\{1,\cdots,N\}$, and the following results are obtained.
\begin{equation}
\label{feas}
{{\frac{P_{2,0}}{N}}[j-(j-1){\beta}{\gamma}]\le{P_{1,0}}\gamma},j \in \{ 1, \cdots ,N\}
\end{equation}

\emph{Remark 1}: Eq. (\ref{rel_sol}) is the feasible optimal solution for Eq. (\ref{obj_fin}), if

1) $\beta \gamma  \ge 1$: Eq. (\ref{feas}) is satisfied for $j=1$, i.e., ${P_{1,0}}\ge{P_{2,0}}\frac{1}{N\gamma}$;

2) $\beta \gamma  < 1$: Eq. (\ref{feas}) is satisfied for $j=N$, i.e., ${P_{1,0}} \ge {P_{2,0}}\frac{{N - (N - 1)\beta \gamma }}{{N\gamma }}$.

Based on \emph{Remark 1}, we can find the closed form solutions to Eq. (\ref{obj_fin}) with various settings of the system parameters $\beta\gamma$. Note that although normally the overall harvesting efficiency $\beta<1$, $\beta \gamma  \ge 1$ may occur in practice, since $\gamma$ is the ratio of two normalized SNRs. 

\section{Optimal Closed Form Power Allocation for $\beta \gamma  \ge 1$}\label{section3}
In this scenario, as illustrated in Table \ref{tab_1}, when SN transmits with $P_{1,0}$ in the first phase, if RN has sufficient amount of power to match $P_{1,0}\gamma$, SN will harvest $P_{1,0}\beta\gamma$ in the second phase, which is greater than the power consumption in the first phase. Consequently, whether $P_{2,0}$ is sufficient or not to match $P_{1,0}$ is the dominant factor determining the overall throughput.

According to \emph{Remark 1}, at phase $j$, the residual power of RN ${\overline {{P_{2,j}}}}$ will be insufficient to match the residual power $\overline {{P_{1,j}}}$ in SN, when 
\begin{equation}\label{usefullater}
\overline {{P_{1,j}}} > {\overline {{P_{2,j}}}}/[{(N-j+1)\gamma}] 
\end{equation}

Assume RN adopts the \emph{conservative power allocation}, i.e., $\alpha=0$, while SN adopts the \emph{greedy power allocation}, i.e., transmits with all of the residual power in each phase. The insufficiency of RN power will occur no later than phase $k+1$, if Eq. (\ref{eqtable}) in Table \ref{tab_1} is satisfied, i.e., $P_{1,0}>P_{ k }^{th}$.
\begin{equation}\label{def}
\left\{\begin{array}{l}
P_{ k }^{th}=\frac{P_{2,0}}{\gamma [ {\left( {N - k} \right){{\left( {\beta \gamma } \right)}^k} + \mathop \sum \nolimits_{j = 1}^{k} {{\left( {\beta \gamma } \right)}^{j-1}}} ]}\text{, }k\in\{0,\cdots,N-1\}\\
P_{ N }^{th}=0
\end{array}
\right.
\end{equation}

\renewcommand{\arraystretch}{1.5}
\begin{table}[ht]
\caption{Greedy SN and Conservative RN, $\alpha=0$} 
\centering\  
\begin{tabular}{|C|D|} 
\hline                        
SN&DN\\
\hline                  
${P_{1,1}}={P_{1,0}}$&${p_1} = {P_{1,1}}\gamma  < \frac{{{P_{2,0}}}}{N}$\\
\hline 
${P_{1,2}}={P_{1,0}}\beta \gamma$&${p_2} = {P_{1,2}}\gamma  < \frac{{{P_{2,0}} - {p_1}}}{{N - 1}}$\\
\hline 
${P_{1,3}}={P_{1,0}}{(\beta \gamma )^2}$&$ \vdots $\\
\hline 
$ \vdots $&${p_k} = {P_{1,k}}\gamma < \frac{{{P_{2,0}} - \sum\nolimits_{j = 1}^{k - 1} {{p_j}} }}{{N - (k - 1)}}$\\
\hline 
${P_{1,k+1}}={P_{1,0}}{(\beta \gamma )^k}$&\begin{equation}\label{eqtable}
\begin{array}{*{20}{l}}
{p_{k + 1}} = {P_{1,k+1}}\gamma \ge \frac{{{P_{2,0}} - \sum\nolimits_{j = 1}^k {{p_{j}}} }}{{N - k}}
\end{array}
\end{equation}\\
\hline 
${P_{1,j}}={P_{1,0}}{(\beta \gamma )^k}$&${p_j} = {p_{k + 1}}$, $j\in\{k+2,\cdots,N\}$\\
\hline 
\end{tabular}
\label{tab_1}
\end{table}

\emph{Proposition 3}: With $\beta \gamma  \ge 1$ and $P_{1,0}$ is between $\left( {P_{k - 1}^{th}},P_{k }^{th} \right]$, $k \in \left\{ {1, \cdots ,N} \right\}$, there exists an $l$, $\left( {l \le k} \right)$, such that the following power allocation strategy can guarantee the maximum throughput
\[\left\{ \begin{array}{l}
{P_{1,j}}={\overline{P_{1,j}}},j\in\{1,\cdots,l\}\\
{P_{2,j}}=\frac{\overline{P_{2,l+1}}}{N-l},j\in\{l+1,\cdots,N\}
\end{array}\right.\]

\begin{proof}
a) From the SN's perspective, the residual power of phase $j+1$ is
\[\overline{P_{1,j+1}}=\overline{P_{1,j}}+\beta{P_{2,j}}-P_{1,j},\forall j\in\{1,\cdots,N-1\}\]
Since $\beta{P_{2,j}}-P_{1,j}\ge(\beta\gamma-1)P_{1,j}$, $\overline{P_{1,j+1}}$ is a non-decreasing function of the transmission power ${P_{1,j}}$. So ideally, SN will transmit with all of the residual power in each phase.

b) From the RN's perspective, to maximize the throughput, the power should be distributed as equally as possible among the $N$ phases. According to the water-filling algorithm, if $\overline{P_{1,j}}\gamma$ is less than the average value, RN will transmit with $\overline{P_{1,j}}\gamma$ in phase $j$. Otherwise, when Eq. (\ref{usefullater}) is satisfied, RN will transmit with the same amount of power from phase $j$ to phase $N$.

c) Since $P_{1,0}\in\left( {P_{k - 1}^{th}},P_{k }^{th} \right]$, Eq. (\ref{usefullater}) is satisfied with $j=k+1$ when ${\alpha} = 0$. If ${\alpha} > 0$, there exists an $l$, $l \le k$, such that SN transmits with all of the residual power from phases $1$ to phase $l$, while RN will divide the residual power equally among phase $l + 1$ to phase $N$.
\end{proof}

\subsection{${P_{1,0}} \ge P_{0}^{th}=\frac{P_{2,0}}{N\gamma}$}
According to \emph{Remark 1}, $EC_1$ is satisfied and the relaxed solution in Eq. (\ref{rel_sol}) is optimal.
\begin{equation}
\label{case1_1}
\left\{ \begin{array}{l}
{P_{2,j}} = {p_j} = \frac{{P_{2,0}}}{N},j \in \{ 1, \cdots ,N\} \\
{P_{1,j}} = \frac{p_j}{\gamma}  = \frac{{P_{2,0}}}{N\gamma},j \in \{ 1, \cdots ,N\} 
\end{array}
\right.
\end{equation}

\subsection{$P_{{k - 1}}^{th} > {P_{1,0}} \ge P_{k}^{th}$, $k \in \left\{ {1, \cdots ,N} \right\}$}
For each $l \in \left\{ {1, \cdots ,k} \right\}$, \emph{Proposition 3} indicates that $p_j$, $j\in\{2,\cdots,N\}$, in Eq. (\ref{obj_fin}) are all determined by $\alpha$, as shown in Table \ref{tab_2}. The feasible domain of $\alpha$ for each $l$ is given by the threshold $\alpha _{l}^{th}$, which is defined as the minimum positive value such that Eq. (\ref{eqtable2}) is satisfied with equality.
\[\left\{ \begin{array}{l}
\alpha _{l}^{th} = {\left\{ { \frac{P_{2,0}}{\left( {N - l} \right){{\left( {\beta \gamma } \right)}^l} + \mathop \sum \nolimits_{j = 1}^l {{\left( {\beta \gamma } \right)}^{j - 1}}}- {P_{1,0}}\gamma} \right\}^ + }\\
\alpha_{0}^{th}=P_{2,0}
\end{array}\right.\]
where $\{\bullet \}^+=max\{\bullet ,0\}$.

\begin{table}[ht]
\caption{Greedy SN and Cooperative RN, $\alpha\ge0$} 
\centering\  
\begin{tabular}{|E|F|}
\hline                        
SN&DN\\
\hline                  
${P_{1,0}}$&${P_{2,1}} = p_1+\alpha={P_{1,0}}{\gamma}  + {\alpha} < \frac{P_{2,0}}{N}$\\
\hline 
$({P_{1,0}}\gamma  + {\alpha})\beta $&$\vdots$\\
\hline 
$ \vdots $&
${P_{2,l}}=p_l=({P_{1,0}}\gamma  + {\alpha}){(\beta \gamma )^{l - 1}} < \frac{{P_{2,0}} - \sum\nolimits_{j = 1}^{l - 1} {{P_{2,j}} }}{{N - (l - 1)}}$\\
\hline 
$({P_{1,0}}\gamma  + {\alpha}){(\beta \gamma )^{l - 1}}\beta$ &
\begin{equation}\label{eqtable2}
{P_{2,l + 1}}= p_{l+1}=({P_{1,0}}\gamma  + {\alpha}){(\beta \gamma )^l} \ge \frac{{P_{2,0}} - \sum\nolimits_{j = 1}^l {{P_{2,j} }}}{{N - l}}
\end{equation}\\
\hline 
\end{tabular}
\label{tab_2}
\end{table}
\begin{figure*}[width=\textwidth]
\hrule
\begin{equation}\label{case1_sim_E}
\begin{array}{l}
\mathop {\max }\limits_{\{ {\alpha}\} } \sum\limits_{j = 2}^l {\log [1 + ({P_{1,0}}\gamma  + {\alpha}){{(\beta \gamma )}^{j - 1}}{\gamma _2}]}  + (N - l)\log \{ 1 + \frac{{[{P_{2,0}} - ({P_{1,0}}\gamma  + {\alpha})\sum\limits_{j = 1}^l {{{(\beta \gamma )}^{j - 1}}} ]{\gamma _2}}}{({{N - l}})}\} \\
{s.t.}\:{\alpha ^{th}_{l - 1}} > {\alpha} \ge {\alpha ^{th}_{l}}
\end{array}
\end{equation}
\hrule
\end{figure*}
\begin{figure*}[width=\textwidth]
\begin{equation}\label{eqcase11}
\left\{\begin{array}{*{20}{l}}
{-\sum\limits_{j = 2}^l {\frac{{{{(\beta \gamma )}^{t - 1}}(1 - \beta \gamma )}}{{1 + ({P_{1,0}}\gamma  + {\alpha}){{(\beta \gamma )}^{j - 1}}{\gamma _2}}}}  +\frac{(N-l)[{1 - {{(\beta \gamma )}^{l - 1}}}]}{{N - l + [{P_{2,0}} - ({P_{1,0}}\gamma  + {\alpha})\sum\limits_{j = 1}^l {{{(\beta \gamma )}^{j - 1}}} ]{\gamma _2}}}=0},
&{\beta\gamma>1}\\
{-{\frac{l-1}{{1 + ({P_{1,0}}\gamma  + {\alpha}){\gamma _2}}}}+\frac{(N-l)l}{{N - l + [{P_{2,0}} - ({P_{1,0}}\gamma  + {\alpha})l ]{\gamma _2}}}=0},
&{\beta\gamma=1}
\end{array}
\right.
\end{equation}
\hrule
\end{figure*}

When ${\alpha ^{th}_{l - 1}} > {\alpha} \ge {\alpha ^{th}_{l}}$, the insufficiency of RN's power will occur at phase $l+1$. As illustrated in Eq. (\ref{case1_sim_E}), the simplified form of Eq. (\ref{obj_fin}) allows us to use the Lagrange method to get the optimal solution by setting the first derivative of the objective function in Eq. (\ref{case1_sim_E}) to zero, as shown in Eq. (\ref{eqcase11}). The corresponding optimal solution is
\begin{equation}
\label{solutioncaseA1}
\left\{ \begin{array}{l}
{P_{2,1}} = {p_1} + {\alpha^*} = {P_{1,0}}\gamma  + {\alpha^*}\\
{P_{2,j}} = ({P_{1,0}}\gamma  + {\alpha^*}){(\beta \gamma )^j},j \in \{ 2, \cdots ,l\} \\
{P_{2,j}} = \frac{({P_{2,0}} - \sum\limits_{t = 1}^l {{P_{2,t}}} )}{N - l},j \in \{ l + 1, \cdots ,N\} \\
{P_{1,1}} = {P_{1,0}},{P_{1,j}} = \frac{P_{2,j}}{\gamma} ,j \in \{ 2, \cdots ,N\} 
\end{array}
\right.
\end{equation}
where ${\alpha^*}$ is the solution to Eq. (\ref{eqcase11}), if it falls within $({\alpha ^{th}_{l - 1}},{\alpha ^{th}_{l }}]$. Otherwise, ${\alpha^*}={\alpha ^{th}_l}$.

\emph{Note:} For the case with $P_{0}^{th} > {P_{1,0}} \ge P_{1}^{th}$, the optimal solution is $\alpha^*=0$, since $k=l=1$ and $\alpha _{l}^{th}=0$.

\section{Optimal Closed Form Power Allocation for $\beta \gamma  < 1$}\label{section4}
Unlike the scenario where SN can rely solely on the harvested energy after the first phase, SN needs to spare part of the initial power storage $P_{1,0}$ for future data transmission with $\beta\gamma<1$. From the throughput's perspective, as compared with $P_{i,j}>P_{i,j+1}$, $P^*_{i,j}=P^*_{i,j+1}$ is always a preferable solution (\emph{Remark 1}). However, $P^*_{i,j}=P^*_{i,j+1}$ may not be feasible. The reason is that from the energy's point of view, $P^*_{i,j}\le{P^*_{i,j+1}}$ will bring less harvested energy to phase $j+1$ than $P_{i,j}>{P_{i,j+1}}$.

To untangle the above mentioned relationship between energy and throughput, we define the \emph{\bf{partial residual power}}, i.e., part of the residual power at the beginning of phase $j-1$ that is used for data transmission in the two consecutive phases $j - 1$ and $j$, as $\overline {{P_{1,j-1}}}'$, $j\in\{2,\cdots,N\}$.
\begin{equation}
\label{partial}
\overline {{P_{1,j - 1}}}'  = {P_{1,j - 1}} + {P_{1,j}} - \beta '{P_{2,j - 1}}
\end{equation}
where ${P_{2,j - 1}} = {\alpha _{j - 1}} + {p_{j - 1}} = {\alpha _{j - 1}} + \gamma {P_{1,j - 1}}$, and $\beta'{P_{2,j-1}},  (0\le\beta'\le\beta)$ is the power harvested in phase $j - 1$ and used in phase $j$.

When $\overline {{P_{1,j - 1}}}' \gamma  < {P_{2,j}}$, the following result can be obtained from Eq. (\ref{partial}) because ${P_{2,j}}= \gamma {P_{1,j}}$, $j\ge2$.
\begin{equation}
\label{eqaaa}
{\alpha _{j - 1}} > \frac{{\left( {1 - \beta '\gamma } \right){P_{1,j - 1}}}}{{\beta '}} > 0
\end{equation}

Then, we can deduce that $\overline {{P_{1,j - 1}}}' \gamma  < {P_{2,j}}$ is feasible only when $j = 2$, since in the optimal solution to Eq. (\ref{obj_fin}), only $\alpha_1 \ge0$ (\emph{Proposition 2}).

\emph{Proposition 4}: When $\beta \gamma  < 1$, there exists an optimal solution with
	\[{P_{2,1}} \ge {P_{2,j - 1}} = {P_{2,j}} \ge {P_{2,N}},\forall j \in \{ 3, \cdots ,N - 1\} \]
\begin{proof} 
a) $\overline {{P_{1,1}}}' \gamma  < {P_{2,2}}$: Suppose in the optimal solution $P_{2,1}< P_{2,2}$, then, from Eq. (\ref{partial})-(\ref{eqaaa}), there is

\begin{equation}
\label{xyz}
\left\{ \begin{array}{l}
{P_{1,2}}-{P_{1,1}} =\overline {{P_{1,1}}}' + \beta '(\gamma {P_{1,1}} + {\alpha})-2{P_{1,1}}\\
P_{1,1}<min\{\beta'(\gamma{P_{1,1}}+\alpha),\overline{P_{1,1}}' \}
\end{array}\right.
\end{equation}
\begin{figure}[t]
\centering
\includegraphics[width=5 in]{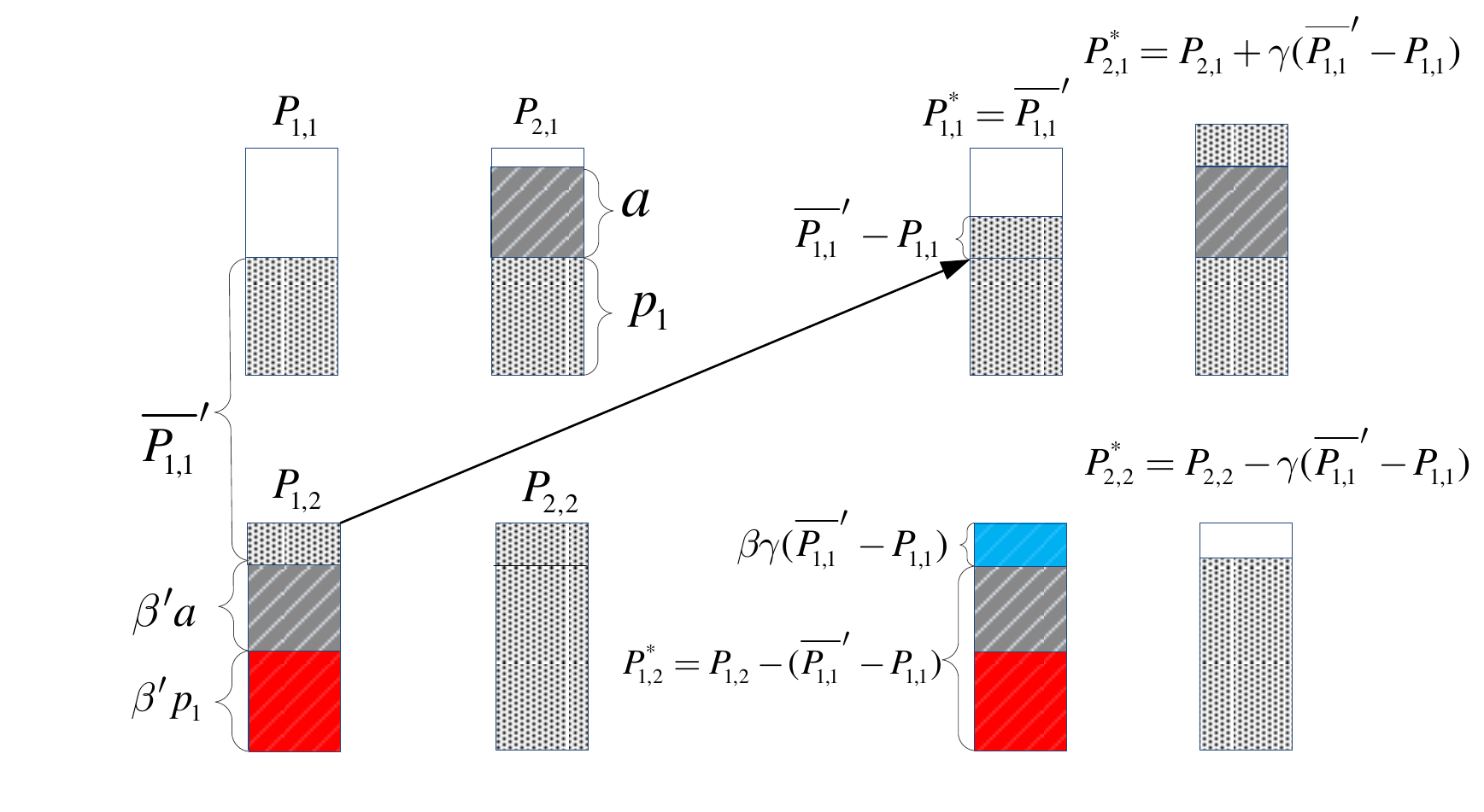}
\caption{$\overline {{P_{1,1}}}' \gamma  < {P_{2,2}}$; the blue area is the increment in harvested energy.}
\label{caseA}
\end{figure}

Consequently, a new feasible allocation with $P^*_{2,1}\ge{P^*_{2,2}}$ exists, as shown in Fig. \ref{caseA}, where
\begin{equation}
\label{xyz1}
\left\{ \begin{array}{l}
P{^*_{1,1}} + P{^*_{1,2}} = {P_{1,1}} + {P_{1,2}}\\
P{^*_{1,2}}-{P{^*_{1,1}}}=  { \beta '(\gamma {P_{1,1}} + {\alpha})}-\overline {{P_{1,1}}}'
\end{array}
\right.
\end{equation}
From Eqs. (\ref{xyz})-(\ref{xyz1}), it can be verified that $\mathop \prod \nolimits_{j=1}^{2} (1+\gamma_1 {P^*_{1,j}}) >\mathop \prod \nolimits_{j=1}^{2}(1+\gamma_1 {P_{1,j}})$, since $\left| {P{^*_{1,1}} - P{^*_{1,2}}} \right|< {{P_{1,2}} - {P_{1,1}}}$. This indicates that the new solution with $P^*_{2,1}\ge{P^*_{2,2}}$ will have greater throughput and harvested energy.

b) $\overline {{P_{1,1}}}' \gamma  \ge {P_{2,2}}$: Similarly to a), suppose the solution with $P_{2,1}< P_{2,2}$ is optimal. Then, 
a new feasible allocation with $P^*_{2,1}\ge{P^*_{2,2}}$ will have an equivalent objective value of sum throughput and greater harvested energy.
\begin{equation}
\label{similar}
\left\{ \begin{array}{l}
P{^*_{1,1}} = {P_{1,2}}$, $P{^*_{1,2}} = {P_{1,1}}\\
P{^*_{2,1}} = {P_{2,2}}+{\alpha}$, $P{^*_{2,2}} = {P_{2,1}} - {\alpha}
\end{array}
\right.
\end{equation}
where the increment in the harvested energy for data transmission in phase $3$ to phase $N$ is $\beta({P^*_{2,1}}-{P_{2,1}})$.

According to a) and b), ${P_{2,1}} \ge {P_{2,2}}$.

c) $\overline {{P_{1,j - 1}}}' \gamma  \ge {P_{2,j}},j>2$: 	Note that when $j > 2$, $\overline {{P_{1,j - 1}}}' \gamma  \ge {P_{2,j}}$ is always true, since ${\alpha _{j - 1}} = 0$ does not satisfy the condition in Eq. (\ref{eqaaa}). Furthermore, simply by setting $\alpha$ in Eq. (\ref{similar}) 
to zero, we can prove that ${P_{2,j - 1}} \ge {P_{2,j}}$, $\forall j \in \left\{ {3, \ldots ,N} \right\}$.

Suppose in the optimal solution, $\exists j \in \{ 3, \cdots ,N - 1\} $ such that $P_{2,j-1}>{P_{2,j}}$, then, depending on $\overline {{P_{1,j-1}}}$, SN's residual power at the $(j-1)$-th phase, new solutions can always be found that will increase the aggregate throughput of the system.

c.1) $\overline {{P_{1,j - 1}}}  \ge \frac{2-{\beta \gamma }}{2\gamma}({{P_{2,j - 1}} + {P_{2,j}}})$: As shown in Fig. \ref{caseC1}, SN will shift $\Delta {P_{1,j - 1}}$ amount of transmission power from phase $(j-1)$ to the $j$-th phase. Meanwhile, to compensate for the harvested energy loss in phase $j$, i.e., $\Delta {P_{1,j - 1}}\beta '\gamma $, the subsequent phases (phase $j+1$ to phase $N$) will decrease harvested energy usage from $(\beta-\beta')$ to $(\beta-\beta'-\Delta{\beta})$, and lessen the residual power usage by $\Delta{\overline {{P_{1,j - 1}}}}$, until the following equalities are satisfied.
\begin{equation}
\label{condition1}
\left\{ \begin{array}{l}
P{^*_{i,j - 1}} = P{^*_{i,j}} = \frac{{{P_{i,j - 1}} + {P_{i,j}}}}{2},i\in\{1,2\}\\
\overline{P^*_{i,j+1}}=\overline{P_{i,j+1}},i\in\{1,2\}
\end{array}
\right.
\end{equation}
where $\overline{P^*_{i,j+1}}$ is the residual power of SN ($i=1$) and RN ($i=2$) with new power allocation.

\begin{figure*}[t]
\center
\includegraphics[width=5in]{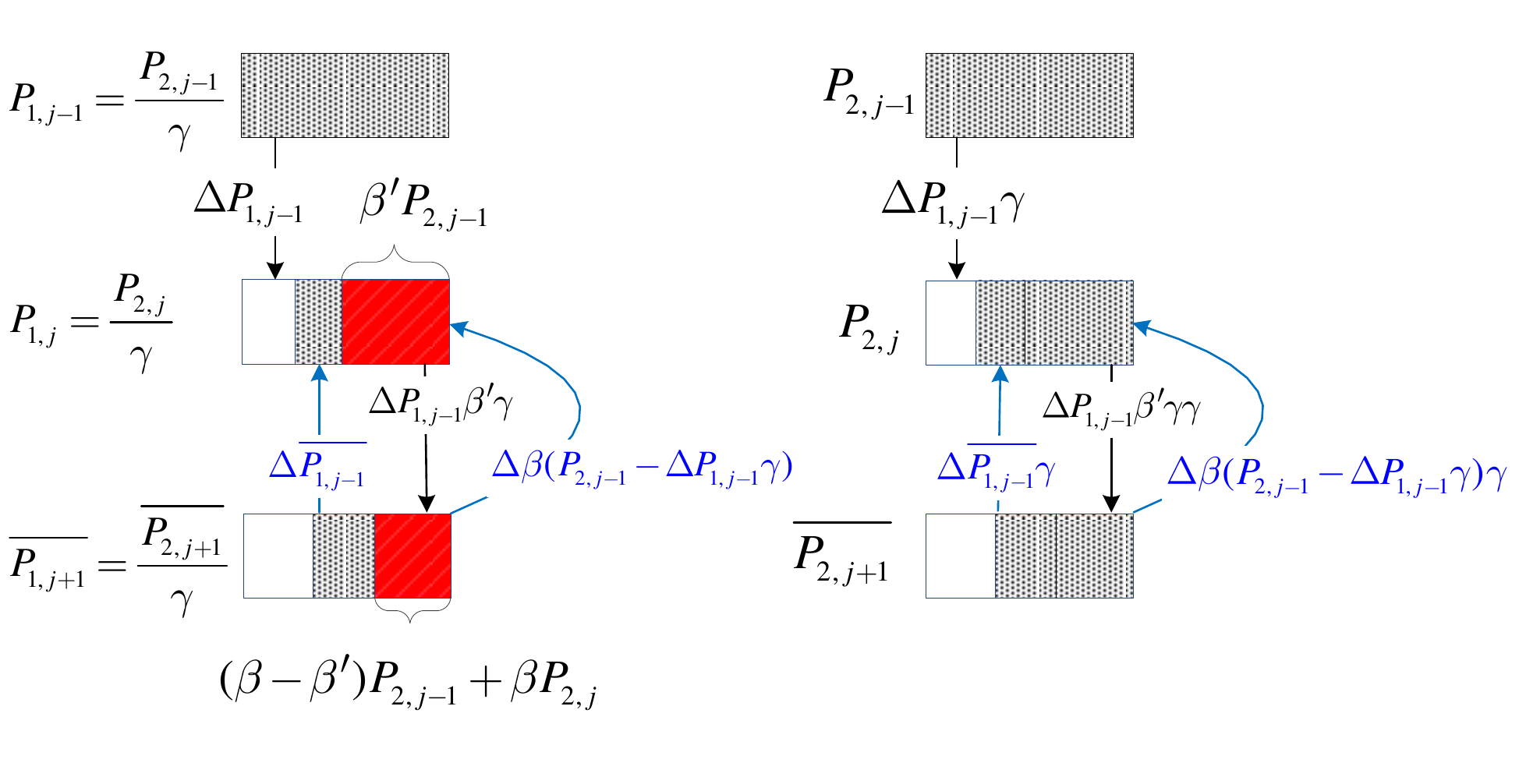}
\caption{$\overline {{P_{1,j - 1}}}' \gamma  \ge {P_{2,j}}$ and $P_{2,j-1}>P_{2,j}$, $j>2$.}
\label{caseC1}
\end{figure*}
c.2) $\overline {{P_{1,j - 1}}} < \frac{2-{\beta \gamma }}{2\gamma}({{P_{2,j - 1}} + {P_{2,j}}})$: The solution with $\mathop{\min }\limits_{i=1,2}\{\overline {P_{i,j+1}}\gamma_i\}\le\mathop{\min }\limits_{i=1,2}\{\overline {P^*_{i,j+1}}\gamma'_i\} \le P{^*_{i,j}}\gamma_i$ is feasible.
\[\left\{ \begin{array}{l}
P{^*_{1,j - 1}} = P{^*_{1,j}} = \frac{{\overline {{P_{1,j - 1}}}}}{{2 - \beta \gamma }}\\
P{^*_{2,j - 1}} = P{^*_{2,j}} = \frac{{\overline {{P_{1,j - 1}}} \gamma }}{{2 - \beta \gamma }}
\end{array}\right.\]
So, the sum throughput from phase $j-1$ to phase $N$ will increase, although the sum throughput of phase $j-1$ and $j$ may not increase.

From c), we can see that ${P_{2,j - 1}} = {P_{2,j}} \ge {P_{2,N}},\forall j \in \{ 3, \cdots ,N - 1\} $. 
\end{proof}

\emph{Remark 2}: There exists ${P_{2,N-1}} \ge {P_{2,N}}$. Furthermore, if ${P_{2,N-1}} > {P_{2,N}}$, there must be ${P_{2,N}}= {\overline{P_{1,N}}}\gamma$.
\begin{proof}
According to the proof c) of \emph{Proposition 4}, ${P_{2,N-1}} \ge {P_{2,N}}$, where the inequality is only possible in c.2) with $\overline {{P_{1,N - 1}}} < \frac{2-{\beta \gamma }}{2\gamma}({{P_{2,N - 1}} + {P_{2,N}}})$. 

Furthermore, if ${P_{2,N-1}} > {P_{2,N}}$ and ${P_{2,N}}< {\overline{P_{1,N}}}\gamma$, we will have
\[{P_{1,N}}< {\overline{P_{1,N}}}\]

Then, a new power allocation scheme with ${P_{1,N }}\le{P^*_{1,N }}\le{P^*_{1,N - 1}}<{P_{1,N-1 }}$ exists.
\begin{align*}
\left\{ \begin{array}{l}
P{^*_{1,N - 1}}  = P{_{1,N-1}}-\frac{\overline {{P_{1,N}}}-{P_{1,N}}  }{\beta \gamma }\\
P{^*_{1,N}}  = P{_{1,N}}+\frac{\overline {{P_{1,N}}}-{P_{1,N}}  }{\beta \gamma }
\end{array}\right.
\end{align*}

As we can see, the new allocation will have better sum throughput while ${P^*_{1,N}}= {\overline{P^*_{1,N}}}$.
\end{proof}

\emph{Remark 3}: According to the proof c) of \emph{Proposition 4}, there exists ${p_{1}} \le {P_{2,2}}$. Furthermore, if ${p_{1}} < {P_{2,2}}$, there must be ${p_1}= {P_{1,0}}\gamma$. 

\subsection{${P_{1,0}} \ge {P_{2,0}}\frac{{N - (N - 1)\beta \gamma }}{{N\gamma }}$}
According to \emph{Remark 1}, Eq. (\ref{case1_1}) satisfies $EC_N$, and thus the relaxed optimal solution is feasible.
\subsection{${P_{1,0}} < {P_{2,0}}\frac{{N - (N - 1)\beta \gamma }}{{N\gamma }}$}
Let ${p_c}=p_j$, $j\in\{2,\cdots,N-1\}$, then from \emph{Proposal 4} and \emph{Remark 2-3}, the constraints in Eq. (\ref{obj_fin}) become
\begin{equation}
\label{constraints}
\left\{ \begin{array}{l}
{p_1} + \alpha  \ge {p_c} \ge \max\{{p_1},{p_N}\}\\
{p_1},{p_N} \ge 0\\
EC_{j},C_1,C_2,j\in\{1,\cdots,N\}
\end{array} \right. 
\end{equation}
Meanwhile, the optimal solution fall into one of the following four cases.


\subsubsection{${p_1} = {P_{1,0}}\gamma  \le {p_c}$, $p_N=p_c$}
In this case, \emph{Remark 1} indicates $EC_{N}$ is sufficient to represent $EC_j$, $j\in\{2,\cdots,N\}$, and the constraints in Eq. (\ref{constraints}) are simplified as
\begin{equation}
\label{final_1}
\left\{ \begin{array}{l}
{P_{1,0}}\gamma  + \alpha  \ge {p_c} \ge {P_{1,0}}\gamma\\
EC_{N},C_1
\end{array} \right. 
\end{equation}
\subsubsection{${p_1} = {p_c} \le {P_{1,0}}\gamma$, $p_N=p_c$}
Similarly, the constraints become Eq. (\ref{final_2})
\begin{equation}
\label{final_2}
\left\{ \begin{array}{l}
{P_{1,0}}\gamma\ge{p_c}\ge0\\
P_{2,0}\ge\alpha\ge0\\
EC_{N},C_1
\end{array} \right. 
\end{equation}
\subsubsection{${p_1} = {P_{1,0}}\gamma  \le {p_c}$, $p_N=\overline{P_{1,N}}\gamma$} The constraints are as follows:
\begin{equation}
\label{final_3}
\left\{ \begin{array}{l}
{P_{1,0}}\gamma + \alpha  \ge {p_c} \ge \max\{{{P_{1,0}}\gamma},{p_N}\}\\
EC_{N-1},EC_{N},C_1
\end{array} \right. 
\end{equation}
where $EC_{N}$ is satisfied with equality. $EC_{N - 1}$ is satisfied if and only if ${p_N} \ge \beta \gamma {p_c}$.
\subsubsection{${p_1} = {p_c} \le {P_{1,0}}\gamma$, $p_N=\overline{P_{1,N}}\gamma$}
The constraints are the same as Eq. (\ref{final_3}).
\begin{equation}
\label{final_4}
\left\{ \begin{array}{l}
{P_{1,0}}\gamma \ge {p_c} \ge {p_N}\ge \beta\gamma{p_c}\\
{p_c},\alpha\ge0\\
EC_{N} \text{ satisfied with equality},C_1
\end{array} \right. 
\end{equation}

With different constraints, the conversions of Eq. (\ref{obj_fin}) are given in Table \ref{table4}, and the corresponding solutions are as follows:
\begin{equation}
\label{solutionfinal2section}
\left\{{\begin{array}{l}
{P_{2,1}} = {p^*_1} + {\alpha^*_k},k \in \{ 1, \cdots ,4\}\\
{P_{2,N}} = {p^*_N},{P_{2,j}} = {p^*_j},j \in \{ 2, \cdots ,N-1\} \\
{P_{1,1}} = {p^*_1}/\gamma,{P_{1,j}} = {P_{2,j}}/\gamma ,j \in \{ 2, \cdots ,N\}
\end{array}} \right.
\end{equation}
where $\alpha^*_k$, $k\in\{1,2\}$ is equal to Eq. (\ref{case1solutionAPLHA}) and Eq. (\ref{case2solutionAPLHA}), respectively, if they are feasible. Otherwise, $\alpha^*_k$ does not exist. When $k\in\{3,4\}$, $\alpha^*_k$ is equal to Eq. (\ref{case3solutionAPLHA}) or Eq. (\ref{case4solutionAPLHA}), if it falls within $[{\alpha}_{min},{\alpha}_{max}]$; otherwise, $\alpha^*={\alpha}_{min}$ or $\alpha^*={\alpha}_{max}$. For each $k$, ${p^*_j}$, $j\in\{1,\cdots,N\}$ are given in the corresponding constraints.

\renewcommand{\arraystretch}{0.9}
\begin{table*}[ht]
\caption{Optimal Power allocation with $\beta\gamma<1$} 
\label{table4} 
\centering\  
\begin{tabular}{|K|L|}
\hline                        
1) Problem&
$\begin{array}{l}
{C^*_1} = \mathop {\max }\limits_{\{ {\alpha}\} } C = \frac{B}{2}\log (1 + {p_1}{\gamma _2}){(1 + {p_c}{\gamma _2})^{N - 2}}(1 + {p_N}{\gamma _2})\\
\begin{array}{*{20}{l}}
s.t.&{\alpha} \ge {\alpha}_{min}=max\{ \frac{{P_{2,0}} - N{P_{1,0}}\gamma}{N}, {{P_{2,0}} - {P_{1,0}}\gamma}- \frac{{(N - 1)\beta \gamma }{P_{2,0}}}{{N - 1 + \beta \gamma }}  \} \\
&{\alpha} \le{\alpha}_{max}= {{P_{2,0}} - N{P_{1,0}}\gamma} \\
&{p_1} = {P_{1,0}}\gamma,{p_N}={p_c} = \frac{{{P_{2,0}} - ({P_{1,0}}\gamma  + {\alpha})}}{{N - 1}}
\end{array}
\end{array}$
\\[0ex]
\hline 
1) Solution&\begin{equation}\label{case1solutionAPLHA}
{\alpha^*_1} = max\{ \frac{{P_{2,0}} - N{P_{1,0}}\gamma}{N}, {{P_{2,0}} - {P_{1,0}}\gamma}- \frac{{(N - 1)\beta \gamma }{P_{2,0}}}{{N - 1 + \beta \gamma }}  \}
\end{equation}
\\[0ex]
\hline 
2) Problem&
$\begin{array}{l}
{C^*_2} = \mathop {\max }\limits_{\{ {\alpha}\} } C = \frac{B}{2}\log (1 + {p_1}{\gamma _2}){(1 + {p_c}{\gamma _2})^{N - 2}}(1 + {p_N}{\gamma _2})\\
\begin{array}{*{20}{l}}
s.t.&{\alpha} \ge{\alpha}_{min}= max\{0, {{P_{2,0}} - N{P_{1,0}}\gamma}, {{P_{2,0}} - {P_{1,0}}\gamma}- \frac{{(N{P_{2,0}} - {P_{1,0}}\gamma)\beta \gamma }}{{N + \beta \gamma }}  \} \\
&{\alpha} \le {\alpha}_{max}={P_{2,0}}\\
&{p_1}  = {p_N}= {p_c}=\frac{P_{2,0}- \alpha}{N }
\end{array}
\end{array}$
\\\hline 
2) Solution&\begin{equation}\label{case2solutionAPLHA}
{\alpha^*_2} = max\{0, {{P_{2,0}} - N{P_{1,0}}\gamma}, {{P_{2,0}} - {P_{1,0}}\gamma}- \frac{{(N{P_{2,0}} - {P_{1,0}}\gamma)\beta \gamma }}{{N + \beta \gamma }}  \}
\end{equation}\\
\hline
3) Problem&
$\begin{array}{l}
{C^*_3} = \mathop {\max }\limits_{\{ {\alpha}\} } C = \frac{B}{2}\log (1 + {p_1}{\gamma _2}){(1 + {p_c}{\gamma _2})^{N - 2}}(1 + {p_N}{\gamma _2})\\
\begin{array}{*{20}{l}}
s.t.&{\alpha} \ge {\alpha}_{min}=max\{ {P_{2,0}} - {P_{1,0}}\gamma  - \frac{{{P_{2,0}}(N - 2 + \beta \gamma )\beta \gamma }}{{(1 + \beta \gamma )\beta \gamma  + N - 2}},\frac{{{P_{2,0}}}}{{1 + (N - 1)\beta \gamma }} - {P_{1,0}}\gamma \} \\
&{\alpha} \le{\alpha}_{max}= min \{ \frac{{{P_{2,0}} - {P_{1,0}}\gamma [1 + (N - 1)\beta \gamma ]}}{{1 + \beta \gamma }},{P_{2,0}} - {P_{1,0}}\gamma  - \frac{{(N - 1)\beta \gamma }}{{N - 1 + \beta \gamma }}{P_{2,0}}\} \\
&{p_1} = {P_{1,0}}\gamma,{p_N} = {P_{2,0}} + \frac{{{P_{1,0}}\gamma  + {\alpha} - {P_{2,0}}}}{{\beta \gamma }},{p_c} = \left\{ \begin{array}{l}
\frac{{{P_{2,0}} - ({P_{1,0}}\gamma  + {\alpha})(1 + \beta \gamma )}}{{(N - 2)\beta \gamma }},N>2 \\
0,N=2
\end{array}
\right.
\end{array}
\end{array}$
\\\hline 
3) Solution&\begin{equation}\label{case3solutionAPLHA}
{\alpha^*_3} = \frac{{{P_{2,0}}}}{{(N - 1)(1 + \beta \gamma )}} + \frac{{{P_{2,0}}(1 - \beta \gamma )(N - 2)}}{{N - 1}} - \frac{{(\beta \gamma)^2 (N - 2)}}{{(1 + \beta \gamma )(N - 1){\gamma _2}}} - {P_{1,0}}\gamma
\end{equation}\\
\hline
4) Problem&
$\begin{array}{l}
{C^*_4} = \mathop {\max }\limits_{\{ {\alpha}\} } C = \frac{B}{2}\log (1 + {p_1}{\gamma _2}){(1 + {p_c}{\gamma _2})^{N - 2}}(1 + {p_N}{\gamma _2})\\
\begin{array}{*{20}{l}}
s.t.&{\alpha} \ge {\alpha}_{min}=max\{ 0,\frac{{{P_{2,0}} - {P_{1,0}}\gamma [1 + (N - 1)\beta \gamma ]}}{{1 + \beta \gamma }},\frac{{{P_{2,0}}[N - 1 - (N - 2)\beta \gamma ] - {P_{1,0}}\gamma (\beta \gamma  + N - 1)}}{{N - 1 + (1 + \beta \gamma )\beta \gamma }}\} \\
&{\alpha} \le {\alpha}_{max}=min \{ \frac{{{P_{2,0}} - {P_{1,0}}\gamma }}{{1 + \beta \gamma }},{P_{2,0}} - {P_{1,0}}\gamma  - \frac{{\beta \gamma (N{P_{2,0}} - {P_{1,0}}\gamma )}}{{N + \beta \gamma }}\} \\
&{p_1} = {p_c} = \frac{{{P_{2,0}} - {P_{1,0}}\gamma  - {\alpha}(1 + \beta \gamma )}}{{(N - 1)\beta \gamma }},{p_N} = {P_{2,0}} + \frac{{{P_{1,0}}\gamma  + {\alpha} - {P_{2,0}}}}{{\beta \gamma }}
\end{array}
\end{array}$
\\\hline 
4) Solution&\begin{equation}\label{case4solutionAPLHA}
{\alpha^*_4} = \frac{{{\gamma _2}({P_{2,0}} - {P_{1,0}}\gamma )[(1 + \beta\gamma )N - \beta\gamma ] - {P_{2,0}}(N - 1)(1 + \beta\gamma )\beta \gamma {\gamma _2} - {{(\beta \gamma )}^2}(N - 1)}}{{N(1 + \beta\gamma ){\gamma _2}}}
\end{equation}\\[2ex]
\hline
\end{tabular}
\label{tab_3}
\end{table*}

\section{System Model with Direct Link between SN and DN}
To this end, we have solved the throughput maximization problem for the relay system where there is no direct link between SN and DN, due to severe channel attenuation. Here, we will discuss the scenario with direct link between SN and DN, $\gamma_1>\gamma'_1>0$, as illustrated in Fig. \ref{scenario}.

First, the objective function in Eq. (\ref{initial}) becomes 
\begin{equation}
\label{initial2}
\mathop {\max }\limits_{{P_{i,j}}} C = \frac{B}{2}\sum\limits_{j = 1}^N \log (1 + \min\{{P_{1,j}}{\gamma _1},{P_{1,j}}{\gamma' _1}+{P_{2,j}}{\gamma_2}\})
\end{equation}

By classifying the residual power of phase $N$ into: 1) $\overline {{P_{2,N}}} {\gamma _2} \ge\overline {{P_{1,N}}} ({\gamma _1}-{\gamma '_1})$ and 2) $\overline {{P_{2,N}}} {\gamma _2} < \overline {{P_{1,N}}} ({\gamma _1}-{\gamma '_1})$, we can prove \emph{Proposition 1} holds for the scenario with direct link between SN and DN. 

Similarly to Eq. (\ref{reas}), it is reasonable to assume 
\[{P_{1,j}}({{\gamma_1}-{\gamma' _1}})\le{P_{2,j}}{\gamma_2}\]
Then, it is obvious that \emph{Proposition 2} also holds for the direct link case. 

Applying \emph{Propositions 1-2}, the optimization problem corresponding to Eq. (\ref{obj_fin}) becomes
\begin{equation}
\label{obj_fin2}
\begin{array}{l}
{C^*} = \mathop {\max }\limits_{\{ {p_j},{\alpha}\} } C = \frac{B}{2}\sum\limits_{j = 1}^N {\log (1 + {p_j}\frac{\gamma _1}{\gamma^*})} \\
\begin{array}{*{20}{l}}
{s.t.}&{EC}_j,C_1,C_2, j\in\{1,\cdots,N\}
\end{array}
\end{array}
\end{equation}
where $\gamma^*=(\gamma_1-\gamma'_1)/\gamma_2$.

Consequently, simply by substituting $\gamma$ with $\gamma^*$, the analysis in Section \ref{section1} - \ref{section4} is still applicable here.

\section{Numerical Results}
Based on the theoretical analysis, optimal power allocation algorithm (OPT) is given in this section to maximize the system throughput.
\begin{algorithm}[width=6in]
\centering
 \renewcommand{\thealgorithm}{}
\caption{OPT algorithm with $\mathcal{O}(N)$ complexity}
\label{alg1}
\begin{algorithmic}[1]
\STATE $C^*=0$
\IF {$\beta\gamma\ge1$}
       \IF {${P_{1,0}} \ge {P_{2,0}}\frac{{1 }}{{N\gamma }}$}
        \STATE {Calculate $C^*$ according to Eq. (\ref{case1_1})}
\ELSE
        \STATE {Calculate $k$, such that $P_{1,0}\in\left( {P_{k - 1}^{th}},P_{k }^{th} \right]$}
 
        \FOR {$l=1$ to $k$}
               \STATE {Calculate $C$ according to Eq. (\ref{solutioncaseA1})}
               \STATE $C^*=\max\{C^*,C\}$
           \ENDFOR
           \ENDIF

\ELSE
\IF {${P_{1,0}} \ge {P_{2,0}}\frac{{N - (N - 1)\beta \gamma }}{{N\gamma }}$}
\STATE{Calculate $C^*$ according to Eq. (\ref{case1_1})}
\ELSE
\FOR {$k=1$ to $4$}
               \STATE {Calculate $C^*$ according to Eq. (\ref{solutionfinal2section}).}
                              \STATE $C^*=\max\{C^*,C\}$
           \ENDFOR

\ENDIF
\ENDIF
         \RETURN $C^*$
\end{algorithmic}
\end{algorithm}

\begin{figure*}
        \centering
        \begin{subfigure}[b]{2 in}
                \includegraphics[width=\textwidth]{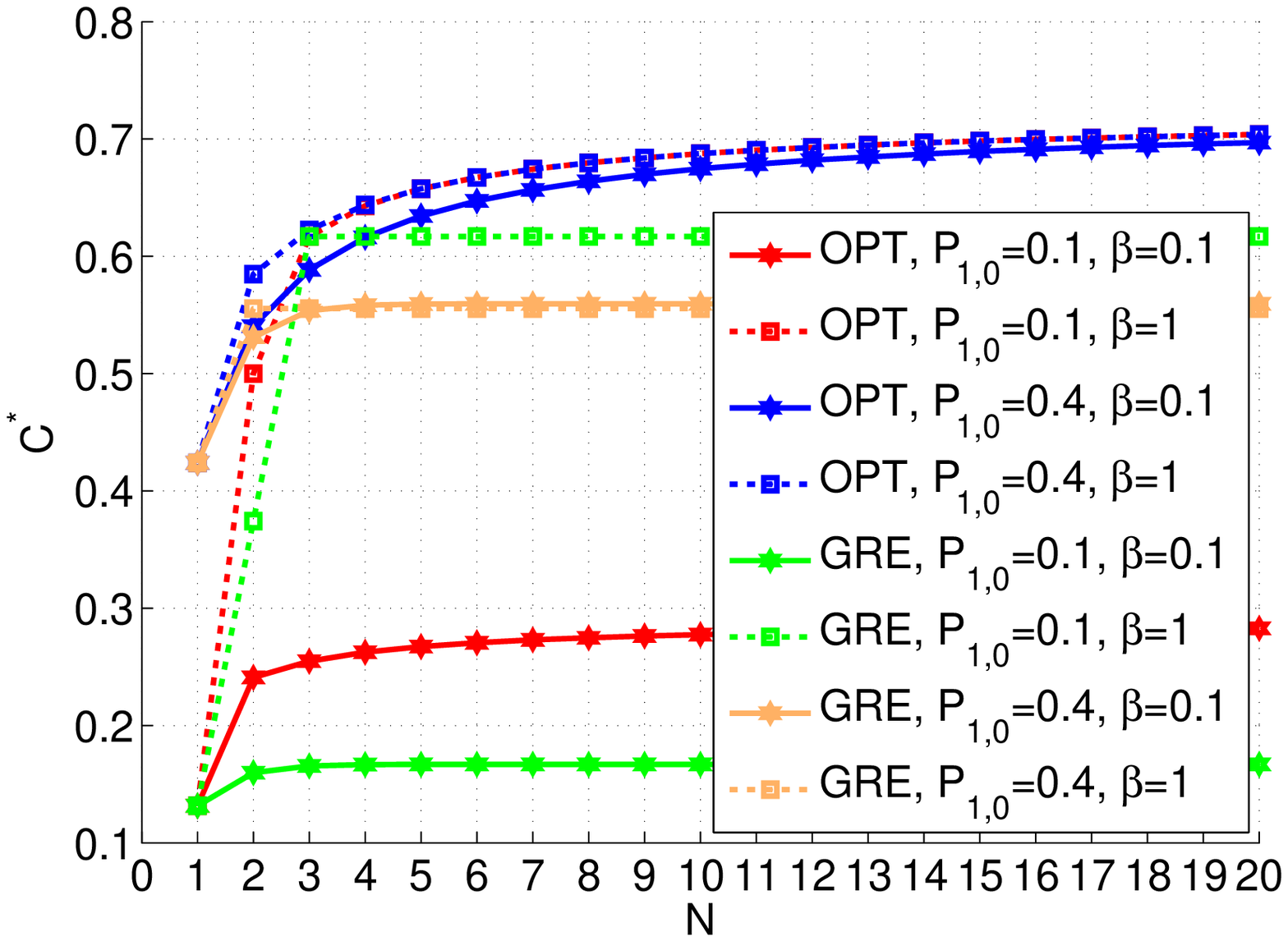}\hspace{0em}%
        \end{subfigure}%
         ~
        \begin{subfigure}[b]{2 in}
                \includegraphics[width=\textwidth]{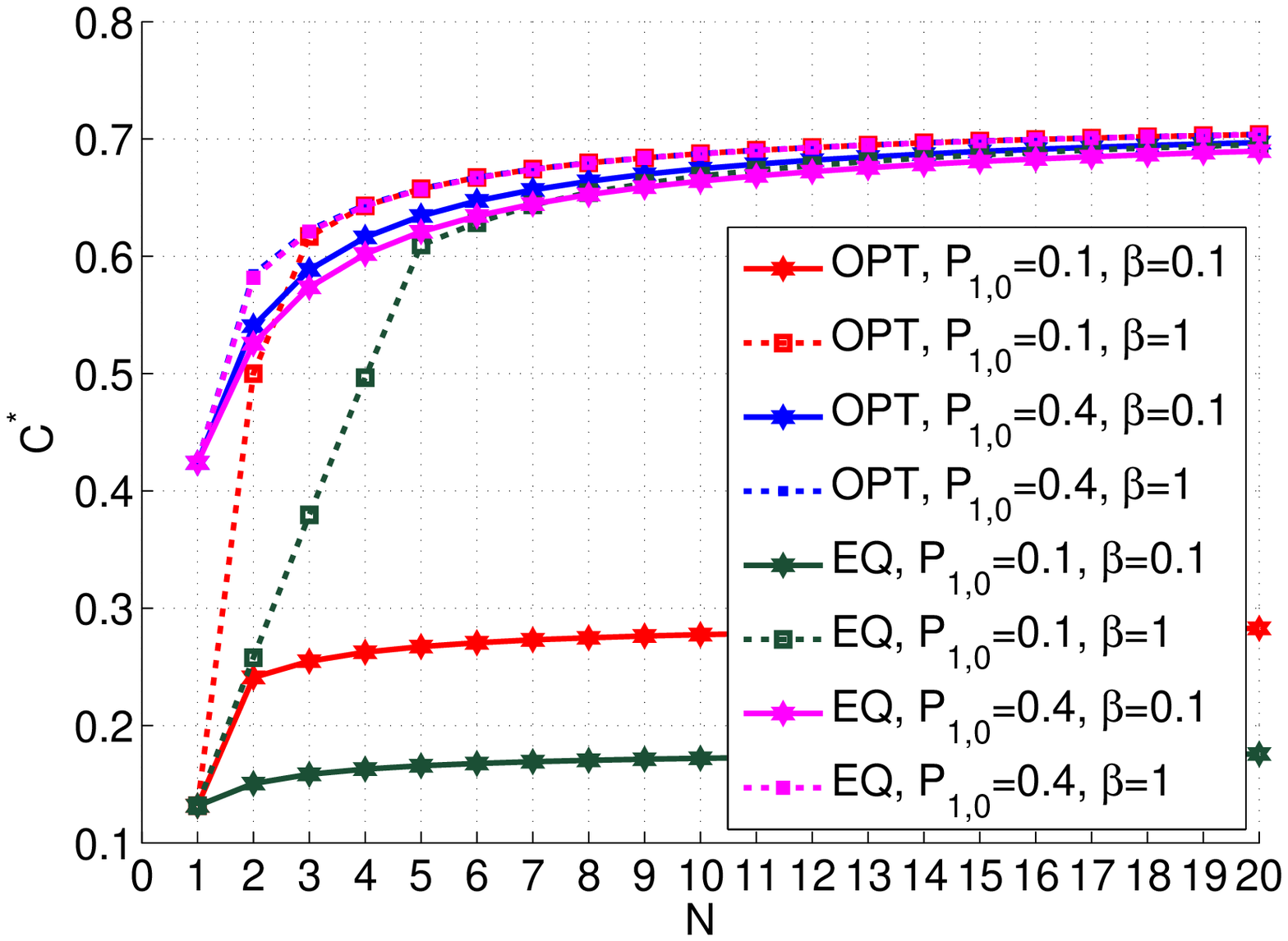}\hspace{0em}%
        \end{subfigure} 
        ~ 
        \begin{subfigure}[b]{2 in}
                \includegraphics[width=\textwidth]{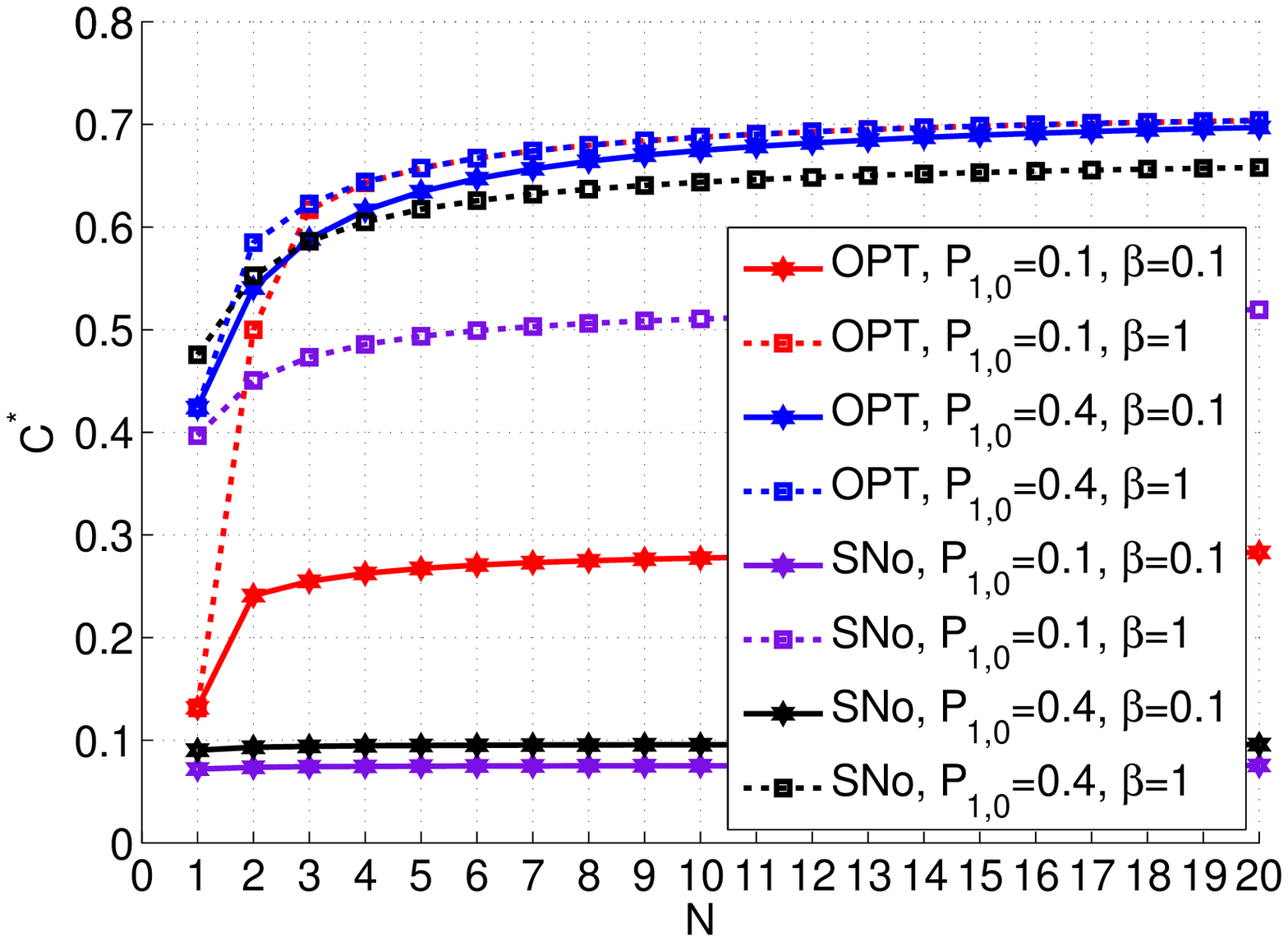}
        \end{subfigure}
        \caption{Comparison of throughput vs. $N$.}\label{CN}
\end{figure*}

\begin{figure*}
        \centering
                \begin{subfigure}[b]{2 in}
                \includegraphics[width=\textwidth]{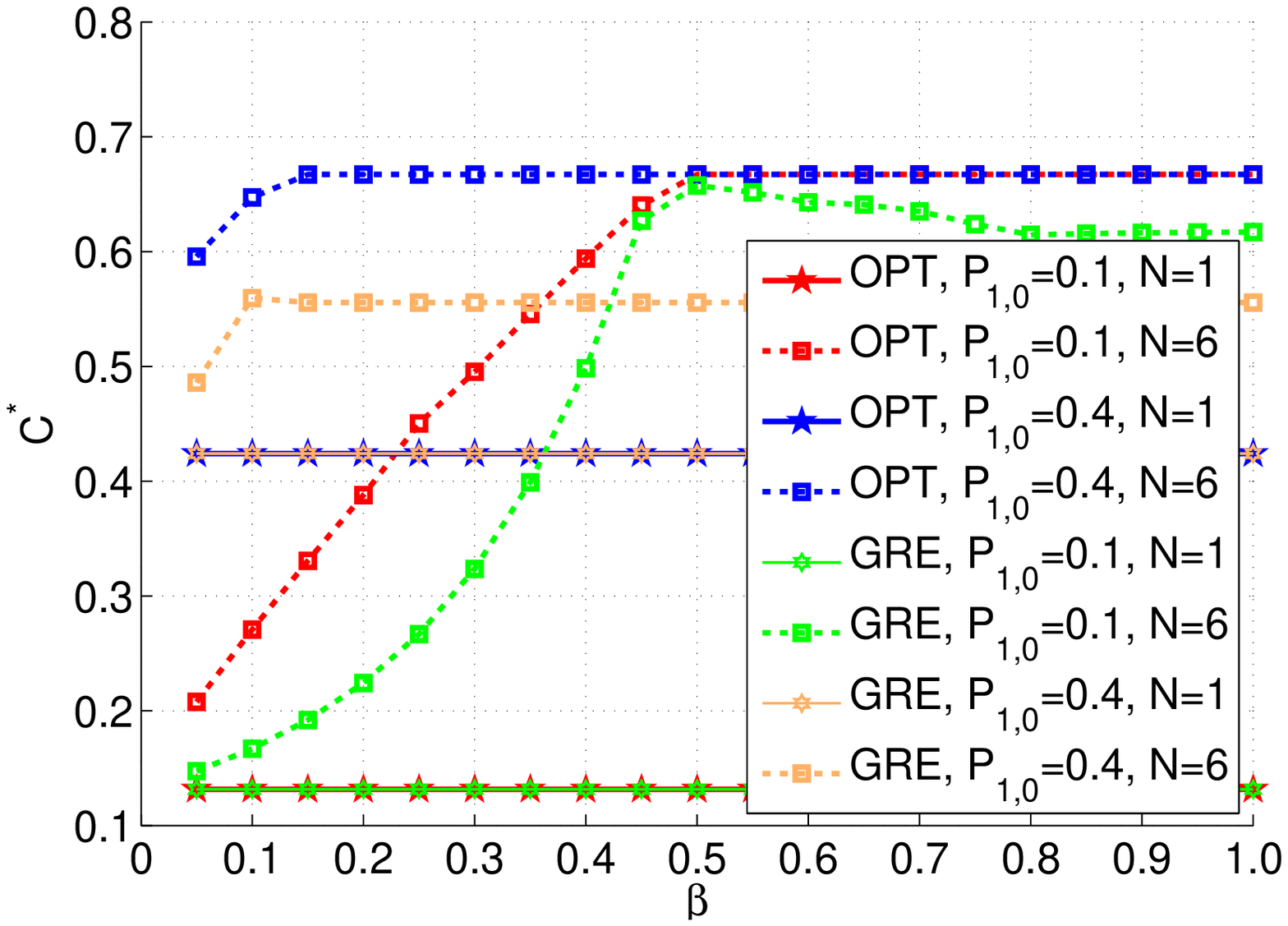}
        \end{subfigure}
        \begin{subfigure}[b]{2 in}
                \includegraphics[width=\textwidth]{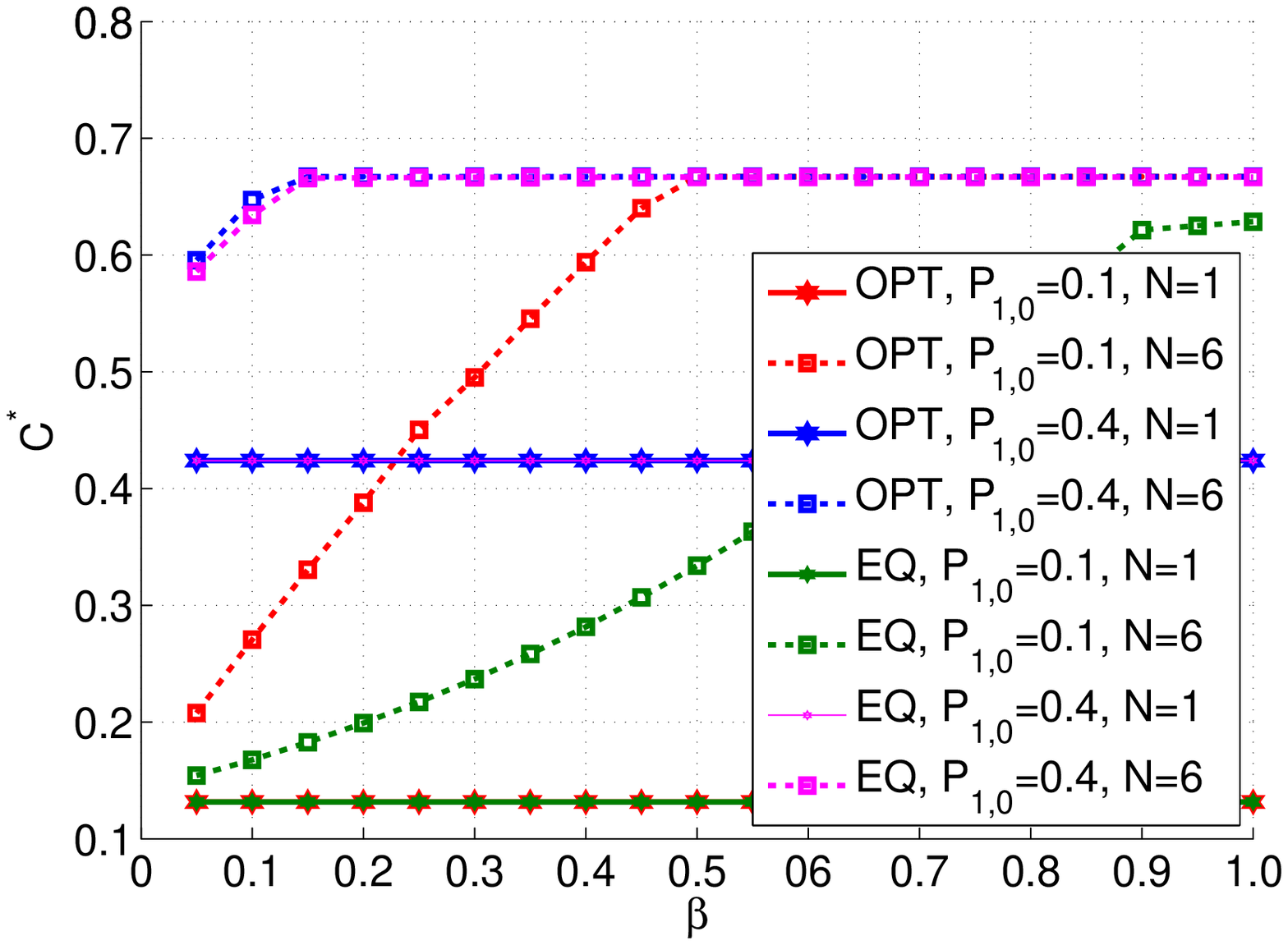}
        \end{subfigure}%
         ~
        \begin{subfigure}[b]{2 in}
                \includegraphics[width=\textwidth]{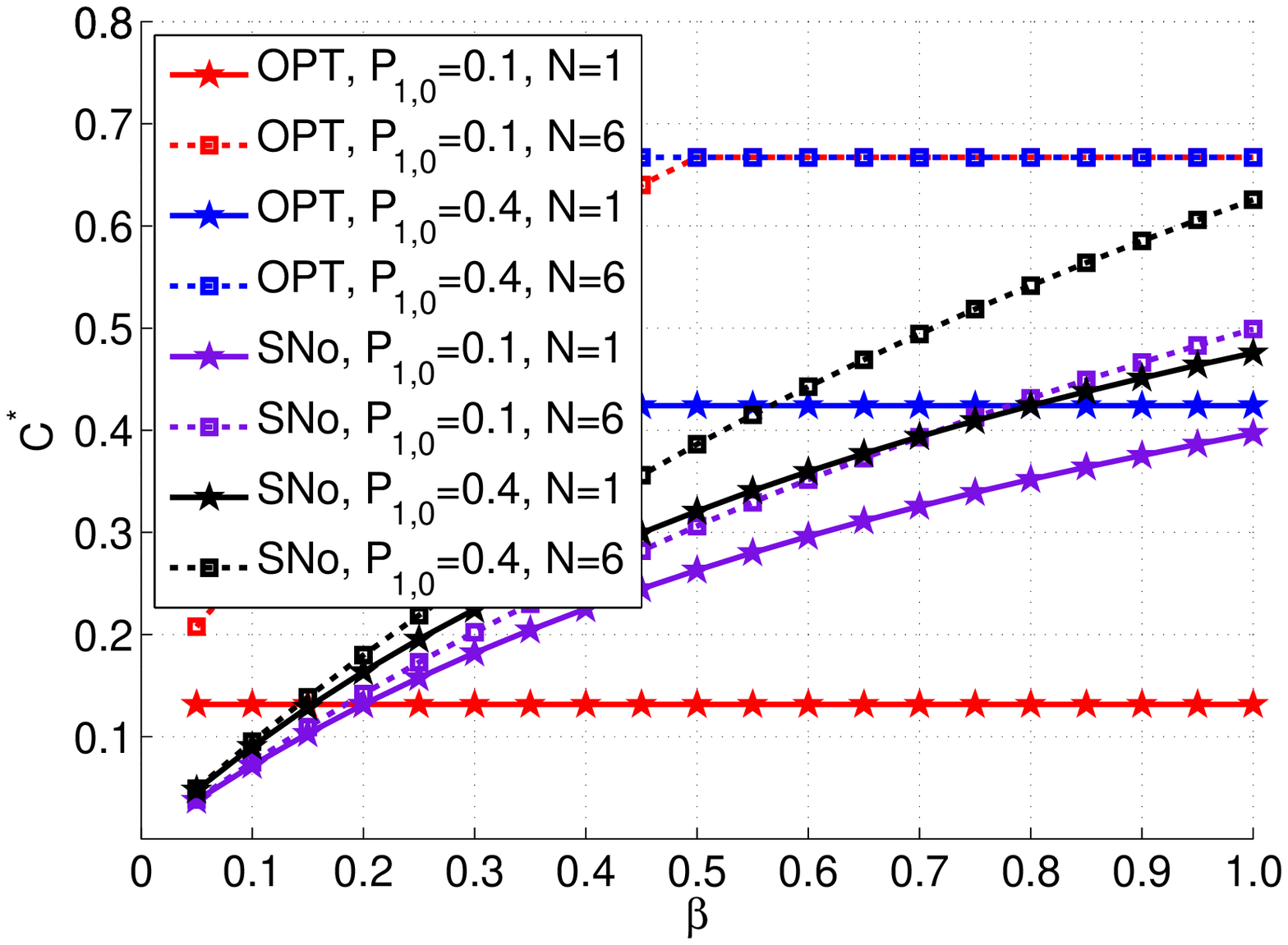}
        \end{subfigure} 
        ~ 

        \caption{Comparison of throughput vs. $\beta$.}\label{BETA}
\end{figure*}

The system has unit bandwidth $B=1$. The budget of reliable power supply for DF-RN is $P_{2,0}=1$, and SNR for the RN-DN link is $\gamma_2=1$. The greedy (GRE), equal (EQ) and SN-only (SNo) power allocation algorithms are used to provide performance reference for our proposed optimal power allocation algorithm. 

In each phase of the GRE algorithm, at least SN or RN will transmit with all of the residual power, depending on the value of $\overline{P_{1,j}}\gamma_1-\overline{P_{2,j}}\gamma_2$. If it is negative, SN will transmit with $\overline{P_{1,j}}$ in phase $j$, and RN will transmit with $\overline{P_{1,j}}\gamma$. If it is positive, $P_{2,j}=\overline{P_{2,j}}$, $P_{1,j}=\overline{P_{2,j}}/\gamma$. Similarly, EQ algorithm states that at least SN or RN will transmit with $\overline{P_{i,j}}/(N-j+1)$, depending on whether $\overline{P_{1,j}}\gamma_1\le\overline{P_{2,j}}\gamma_2$ or not. Finally, SNo algorithm is designed for a system where SN has total power supply of $P_{1,0}+P_{2,0}$, and RN uses the harvested energy to forward data. To maximize throughput, SN will distribute them equally among the $N$ phases and RN has a power split ratio of $\gamma/\beta$, where $\gamma/(\gamma+\beta)$ percent of the received signal is used for energy harvesting, the rest $\beta/(\gamma+\beta)$ percent is used for data detection.
\begin{figure*}
        \centering
        \begin{subfigure}[b]{3 in}
                \includegraphics[width=\textwidth]{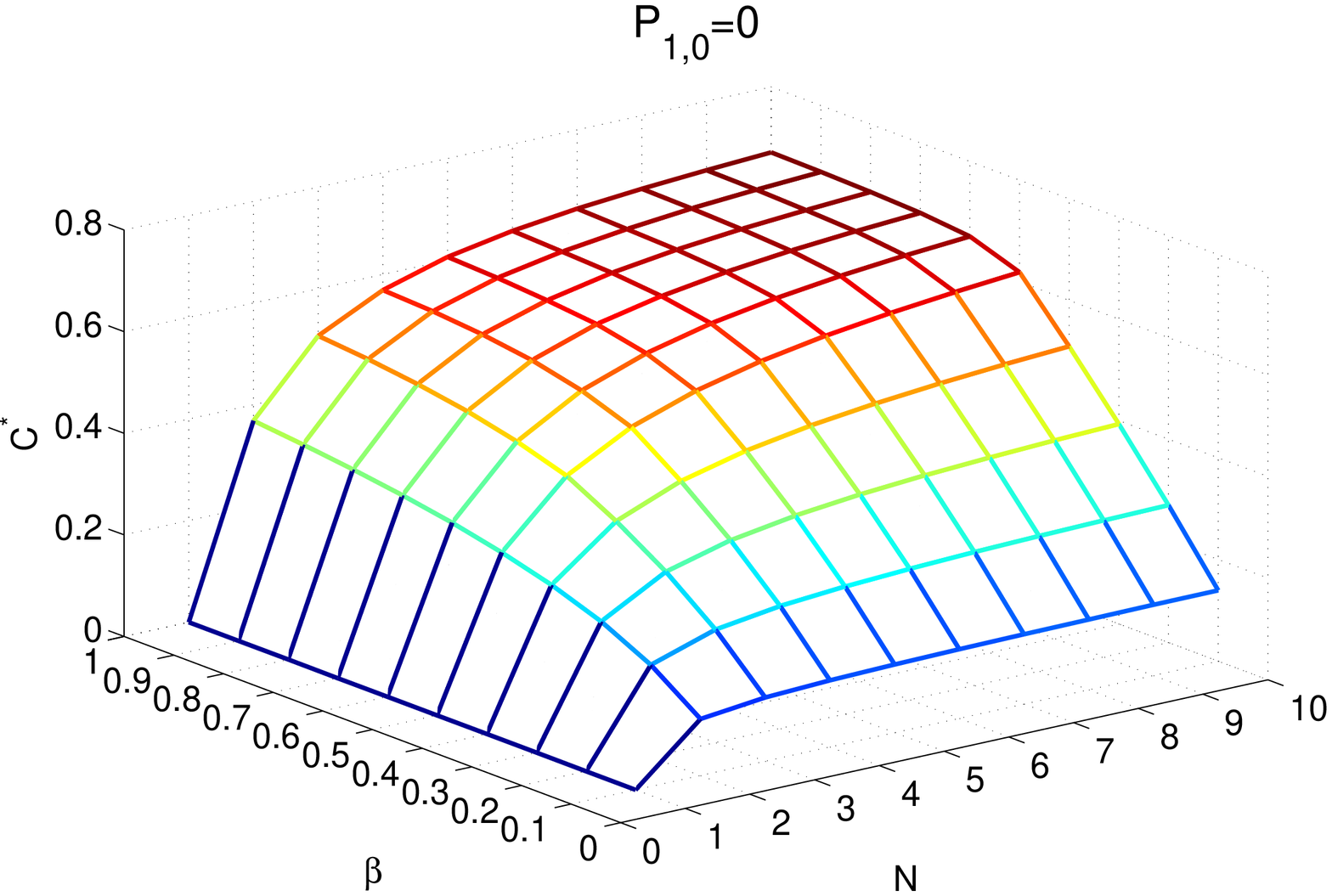}
        \end{subfigure}%
        ~ 
        \begin{subfigure}[b]{3 in}
                \includegraphics[width=\textwidth]{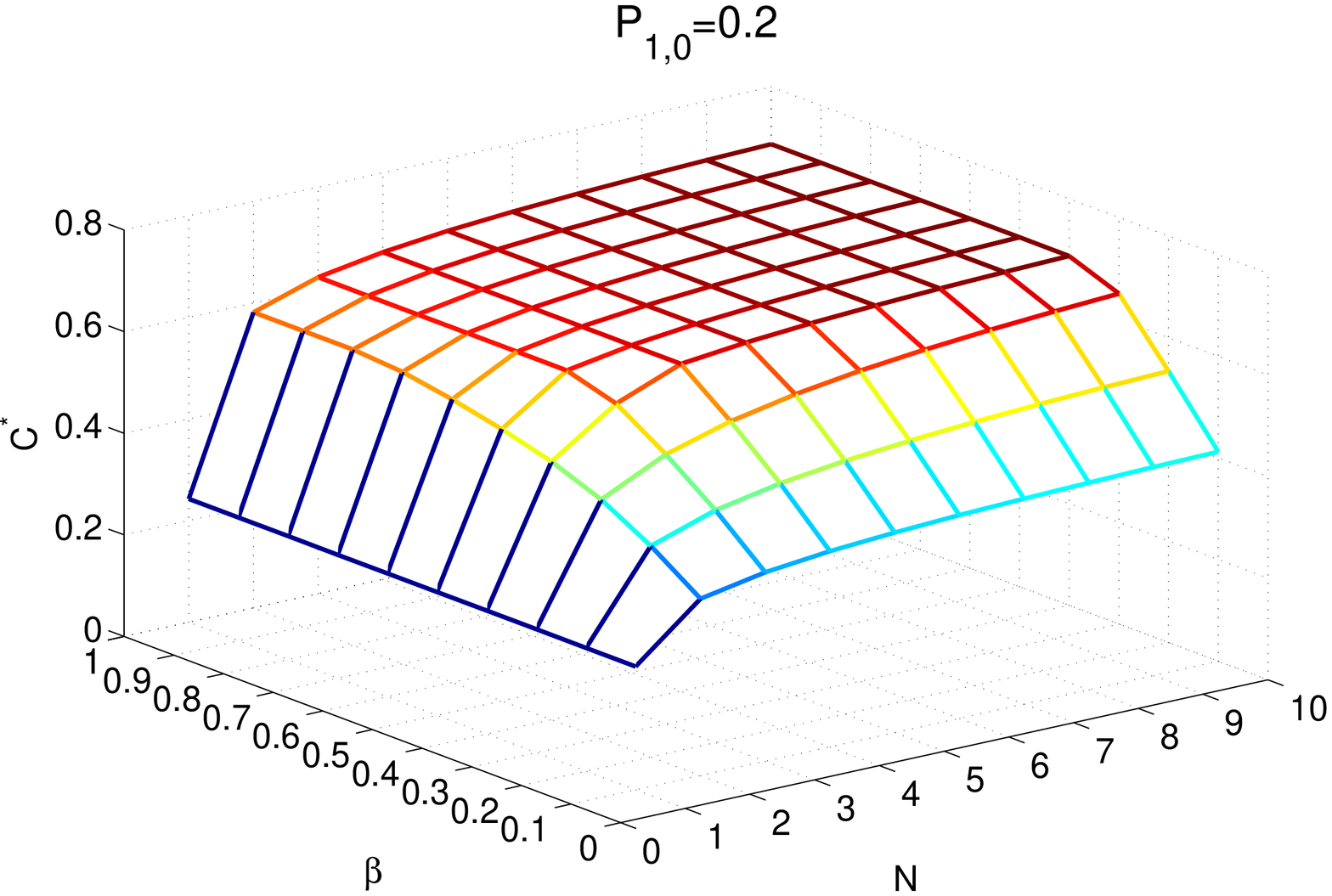}
        \end{subfigure}\\
        \begin{subfigure}[b]{3 in}
                \includegraphics[width=\textwidth]{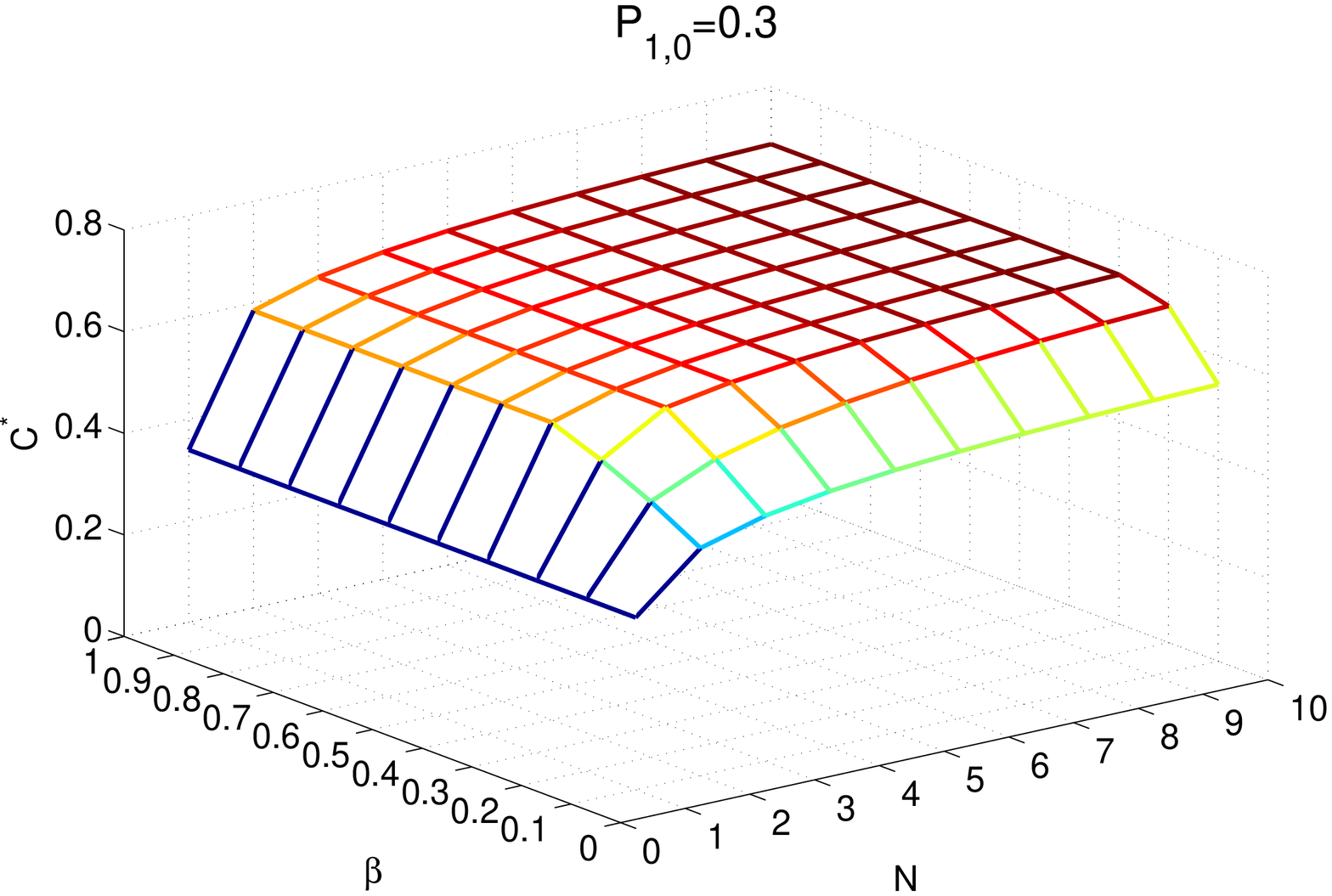}
        \end{subfigure}
        ~
          \begin{subfigure}[b]{3 in}
                \includegraphics[width=\textwidth]{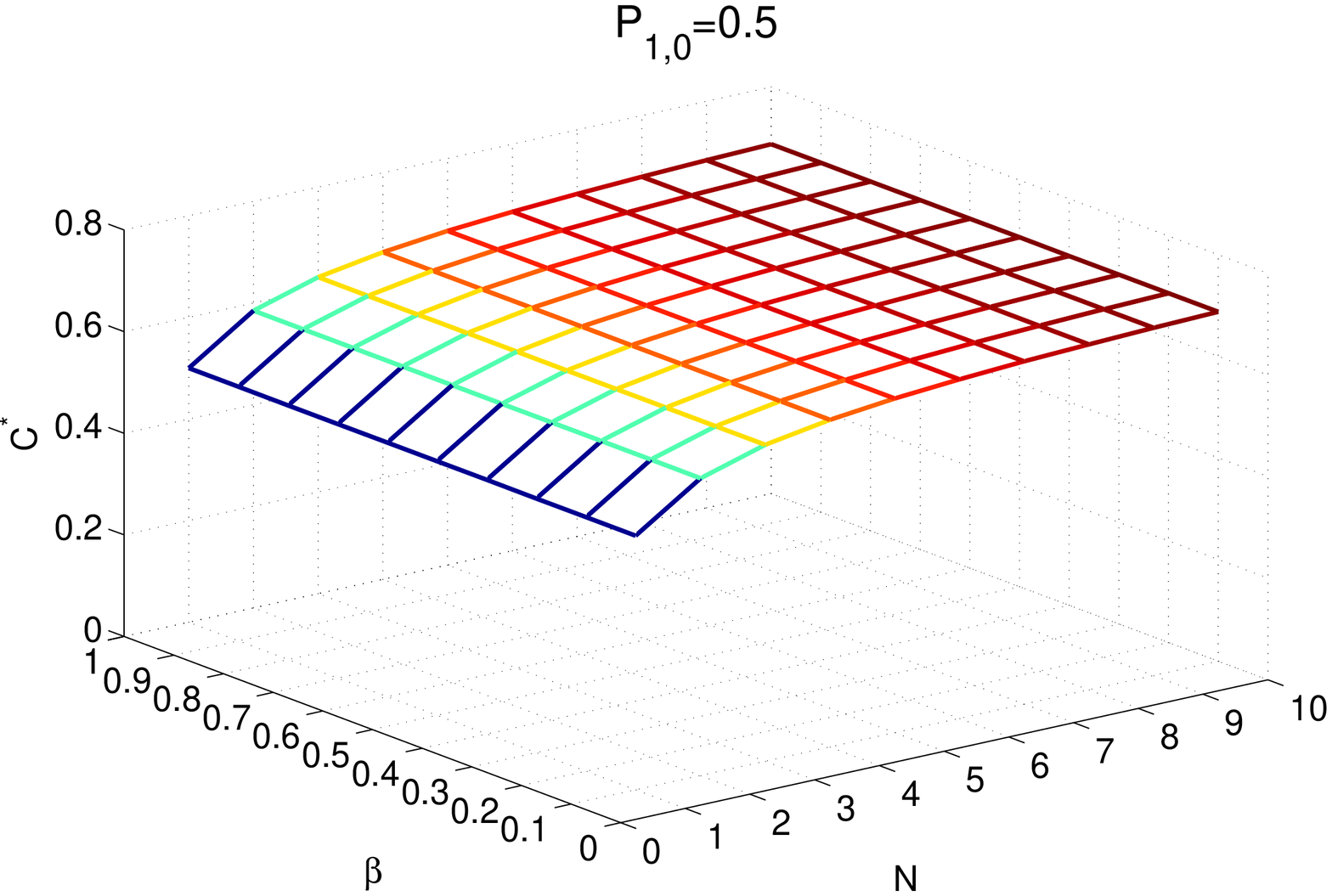}
        \end{subfigure}
        \caption{Optimal throughput with $\gamma_1=2$.}\label{fig:result}
\end{figure*}

Through numerical results shown in Figs. \ref{CN} and \ref{BETA}, the theoretical results are validated. In the OPT algorithm, the system throughput will increase with $N$. More specifically, when $N=1$, the system throughput will only increase with $P_{1,0}$ because the harvested energy cannot be utilized in the first phase. Meanwhile, as $N$ grows, the increment in the overall throughput is less obvious. This is because the total transmission power budget of DN is limited by the on-grid power source, and $P_{2,0}$ will be the leading deciding factor of the overall throughput. Similarly, the system performance will increase with the harvesting efficiency $\beta$, and the performance improvement will be less obvious as $\beta$ increases to a certain point where the power resource of the relay node is more stringent.

Our OPT algorithm always outperforms the GRE algorithm. The performance of OPT and EQ will converge when $\beta$ or $N$ increases. This is because high $\beta$ or $N$ will relax the demanding for $P_{1,0}$ (\emph{Remark 1}), and equal power allocation will become the feasible optimal solution. As compared with SNo algorithm, OPT algorithm will have lower throughput with $N=1$. This is because the energy harvested by SN cannot be used to improve the throughput in the first phase, while the energy harvesting RN (EH-RN) can transmit data using the harnessed energy in the second time slot of the first phase. As we can see, when $N$ increases, the performance difference between OPT and SNo algorithms indicates that in the half duplex relay system, when RN is equipped with EH, SN will not gain anything in the even time slots, while with EH-SN, it can always harvest energy from the signals received in the even TSs.

As illustrated in Fig. \ref{fig:result}, with a certain amount of throughput requirement, instead of demanding high available power from either on-grid power supply, or green energy source, the system can improve energy harvesting efficiency, or utilize more time to transmit delay tolerant traffic using the harvested power.

\section{Conclusion}
Radio frequency energy harvesting provides a new approach for wireless devices to share each other's energy storage, either on-grid power or green power. With simultaneous data and energy transmission, it can also decrease the total power consumption of the wireless system. This is of particular interest to sensor networks where nodes have limited storage capacity, and cellular networks where handsets try to maximize the throughput within time limits. In this paper, we have studied the throughput maximization problem for the orthogonal relay channel with EH-source and relay nodes, assuming a deterministic EH model. For both cases with and without direct link between SN and DN, we have derived the closed form solutions for the optimal joint source and relay power allocation problem. The developed algorithm can achieve the optimal solution for each system setting with linear complexity.

\bibliographystyle{IEEEtran}
\bibliography{mybib}

\begin{thebibliography}{10}
\providecommand{\url}[1]{#1}
\csname url@samestyle\endcsname
\providecommand{\newblock}{\relax}
\providecommand{\bibinfo}[2]{#2}
\providecommand{\BIBentrySTDinterwordspacing}{\spaceskip=0pt\relax}
\providecommand{\BIBentryALTinterwordstretchfactor}{4}
\providecommand{\BIBentryALTinterwordspacing}{\spaceskip=\fontdimen2\font plus
\BIBentryALTinterwordstretchfactor\fontdimen3\font minus
  \fontdimen4\font\relax}
\providecommand{\BIBforeignlanguage}[2]{{%
\expandafter\ifx\csname l@#1\endcsname\relax
\typeout{** WARNING: IEEEtran.bst: No hyphenation pattern has been}%
\typeout{** loaded for the language `#1'. Using the pattern for}%
\typeout{** the default language instead.}%
\else
\language=\csname l@#1\endcsname
\fi
#2}}
\providecommand{\BIBdecl}{\relax}
\BIBdecl

\bibitem{6472203}
T.~Han and N.~Ansari, ``{On greening cellular networks via multicell
  cooperation},'' \emph{Wireless Communications, IEEE}, vol.~20, no.~1, pp.
  82--89, Feb. 2013.

\bibitem{Han:2012:ICE}
{T. Han and N. Ansari}, ``{On Optimizing Green Energy Utilization for Cellular
  Networks with Hybrid Energy Supplies},'' \emph{Wireless Communications, IEEE
  Transactions on}, vol.~12, no.~8, pp. 3872--3882, Aug. 2013.

\bibitem{RF}
T.~Le, K.~Mayaram, and T.~Fiez, ``{Efficient Far-Field Radio Frequency Energy
  Harvesting for Passively Powered Sensor Networks},'' \emph{Solid-State
  Circuits, IEEE Journal of}, vol.~43, no.~5, pp. 1287--1302, May 2008.

\bibitem{RFHARVESTOR}
H.~Jabbar, Y.~Song, and T.~Jeong, ``{RF energy harvesting system and circuits
  for charging of mobile devices},'' \emph{Consumer Electronics, IEEE
  Transactions on}, vol.~56, no.~1, pp. 247--253, Feb. 2010.

\bibitem{6449245}
S.~Luo, R.~Zhang, and T.~J. Lim, ``{Optimal Save-Then-Transmit Protocol for
  Energy Harvesting Wireless Transmitters},'' \emph{Wireless Communications,
  IEEE Transactions on}, vol.~12, no.~3, pp. 1196--1207, Mar. 2013.

\bibitem{5522465}
S.~Sudevalayam and P.~Kulkarni, ``{Energy Harvesting Sensor Nodes: Survey and
  Implications},'' \emph{Communications Surveys Tutorials, IEEE}, vol.~13,
  no.~3, pp. 443--461, Sep. 2011.

\bibitem{energyCausality}
J.~Yang and S.~Ulukus, ``{Optimal Packet Scheduling in an Energy Harvesting
  Communication System},'' \emph{Communications, IEEE Transactions on},
  vol.~60, no.~1, pp. 220--230, Jan. 2012.

\bibitem{6202352}
C.~K. Ho and R.~Zhang, ``{Optimal Energy Allocation for Wireless Communications
  With Energy Harvesting Constraints},'' \emph{Signal Processing, IEEE
  Transactions on}, vol.~60, no.~9, pp. 4808--4818, Sep. 2012.

\bibitem{5441354}
V.~Sharma, U.~Mukherji, V.~Joseph, and S.~Gupta, ``{Optimal energy management
  policies for energy harvesting sensor nodes},'' \emph{Wireless
  Communications, IEEE Transactions on}, vol.~9, no.~4, pp. 1326--1336, Apr.
  2010.

\bibitem{5992841}
O.~Ozel, K.~Tutuncuoglu, J.~Yang, S.~Ulukus, and A.~Yener, ``{Transmission with
  Energy Harvesting Nodes in Fading Wireless Channels: Optimal Policies},''
  \emph{Selected Areas in Communications, IEEE Journal on}, vol.~29, no.~8, pp.
  1732--1743, Sep. 2011.

\bibitem{6363767}
C.~Huang, R.~Zhang, and S.~Cui, ``{Delay-constrained Gaussian relay channel
  with energy harvesting nodes},'' in \emph{Communications (ICC), 2012 IEEE
  International Conference on}, Jun. 2012, pp. 2433--2438.

\bibitem{6381384}
------, ``{Throughput Maximization for the Gaussian Relay Channel with Energy
  Harvesting Constraints},'' \emph{Selected Areas in Communications, IEEE
  Journal on}, vol.~31, no.~8, pp. 1469--1479, Aug. 2013.

\bibitem{6489506}
R.~Zhang and C.~K. Ho, ``{MIMO Broadcasting for Simultaneous Wireless
  Information and Power Transfer},'' \emph{Wireless Communications, IEEE
  Transactions on}, vol.~12, no.~5, pp. 1989--2001, May 2013.

\bibitem{6552840}
A.~Nasir, X.~Zhou, S.~Durrani, and R.~Kennedy, ``{Relaying Protocols for
  Wireless Energy Harvesting and Information Processing},'' \emph{Wireless
  Communications, IEEE Transactions on}, vol.~12, no.~7, pp. 3622--3636, Jul.
  2013.

\bibitem{1435648}
A.~Host-Madsen and J.~Zhang, ``{Capacity bounds and power allocation for
  wireless relay channels},'' \emph{Information Theory, IEEE Transactions on},
  vol.~51, no.~6, pp. 2020--2040, Jun. 2005.

\bibitem{6702854}
Z.~Ding, S.~Perlaza, I.~Esnaola, and H.~Poor, ``{Power Allocation Strategies in
  Energy Harvesting Wireless Cooperative Networks},'' \emph{Wireless
  Communications, IEEE Transactions on}, vol.~PP, no.~99, pp. 1--15, Feb. 2014.

\bibitem{Boyd:2004:CO:993483}
S.~Boyd and L.~Vandenberghe, \emph{{Convex Optimization}}.\hskip 1em plus 0.5em
  minus 0.4em\relax New York, NY, USA: Cambridge University Press, 2004.

\end{thebibliography}

\end{document}